\batchmode
\makeatletter
\def\input@path{{\string"/Users/omuralidharan/Google Drive/Projects/Emp Bayes CI/\string"/}}
\makeatother
\documentclass[english]{amsart}
\usepackage[T1]{fontenc}
\usepackage[latin9]{inputenc}
\usepackage{amsthm}
\usepackage{amsbsy}
\usepackage{amssymb}
\usepackage{graphicx}
\usepackage{esint}
\usepackage[authoryear]{natbib}

\makeatletter
\theoremstyle{plain}
\newtheorem{thm}{\protect\theoremname}
  \theoremstyle{plain}
  \newtheorem{lem}{\protect\lemmaname}

\@ifundefined{showcaptionsetup}{}{%
 \PassOptionsToPackage{caption=false}{subfig}}
\usepackage{subfig}
\makeatother

\usepackage{babel}
  \providecommand{\lemmaname}{Lemma}
\providecommand{\theoremname}{Theorem}

\begin{document}
\global\long\def\P{\mathrm{P}}
\global\long\def\E{\mathrm{E}}
\global\long\def\Var{\mathrm{Var}}
\global\long\def\given{\!\!\bigm|\!\!}
\global\long\def\x{\boldsymbol{x}}
\global\long\def\Poisson{\mathrm{Poisson}}
\global\long\def\Binomial{\mathrm{Binomial}}
\global\long\def\Uniform{\mathrm{Uniform}}
\global\long\def\param{\theta}
\global\long\def\baseestimate{t}
\global\long\def\prior{G_{\baseestimate}}
\global\long\def\priorestimate{\hat{G}_{\baseestimate}}
\global\long\def\conddist{\param\given\baseestimate}
\global\long\def\conddisti{\param_{i}\given\baseestimate_{i}}

\title{Second Order Calibration: A Simple Way to Get Approximate Posteriors}

\author{Omkar Muralidharan and Amir Najmi\\
Google, Inc.}
\begin{abstract}
Many large-scale machine learning problems involve estimating an unknown
parameter $\param_{i}$ for each of many items. For example, a key
problem in sponsored search is to estimate the click through rate
(CTR) of each of billions of query-ad pairs. Most common methods,
though, only give a point estimate of each $\theta_{i}$. A posterior
distribution for each $\param_{i}$ is usually more useful but harder
to get.

We present a simple post-processing technique that takes point estimates
or scores $\baseestimate_{i}$ (from any method) and estimates an
approximate posterior for each $\param_{i}$. We build on the idea
of calibration, a common post-processing technique that estimates
$\E\left(\param_{i}\given\baseestimate_{i}\right)$. Our method, \emph{second
order calibration}, uses empirical Bayes methods to estimate the distribution
of $\conddisti$ and uses the estimated distribution as an approximation
to the posterior distribution of $\param_{i}$. We show that this
can yield improved point estimates and useful accuracy estimates.
The method scales to large problems - our motivating example is a
CTR estimation problem involving tens of billions of query-ad pairs.
\end{abstract}
\maketitle

\section{Introduction\label{sec:Introduction}}

Suppose we have a regression problem: we have responses $y_{i}$ and
covariates $\x_{i}$ for items $i=1,\ldots,I$, and we want to estimate
a parameter $\param_{i}$ for each item using our responses and covariates.
We often want a posterior distribution for each $\param_{i}$, but
common regression methods, like neural networks, boosted trees and
penalized GLMs, only give us point estimates $\baseestimate_{i}=\baseestimate\left(\x_{i}\right)$.
In this paper, we show how to post-process any regression method's
point estimates to get an approximate posterior.

Our motivating problem is click through rate (CTR) estimation for
sponsored search \citep{Richardson2007}. Here, our items are query-ad
pairs, and for each pair, we know the number of times it was shown
(the number of ``impressions'', $N_{i}$), the number of times it
was clicked ($y_{i}$), and covariates that describe the query, ad
and match between them, ($\x_{i}$). We assume the clicks for each
query-ad pair follow a Poisson distribution:
\[
y_{i}\sim\Poisson\left(\param_{i}N_{i}\right)
\]
and want to estimate $\param_{i}$, the CTR for the query-ad pair.
A complex machine learning system gives us point estimates $\baseestimate_{i}$.
We want a posterior distribution for each $\param_{i}$. Among other
things, we could use these posteriors to make fine-grained accuracy
estimates and explore-exploit tradeoffs.

Any method to get posteriors for the CTR problem has to have three
important features. First, it has to scale. Search engines have billions
of query-ad pairs. Second, it cannot depend on the underlying machine
learning system that gives $t_{i}$. CTR estimation systems are complex
- not least because of their scale - and are constantly being improved
\citep{Graepel2010}. If a method used detailed knowledge of the system
to get posteriors, it would have to account for all the complexity
and be updated constantly as the system changed; we also wouldn't
be able to use its posteriors to compare the system to a very different
competitor. Third, the method must share information across query-ad
pairs. Because of the long-tail nature of search queries, most query-ad
pairs are shown a small number of times. This, combined with CTRs
that are generally low, means that taken individually, most query-ad
pairs have little information.

We get approximate posteriors by extending the idea of calibration,
a common post-processing technique that removes bias from regression
estimates. There are a few different methods for calibration, but
all are based the same idea: instead of using $\baseestimate$, estimate
and use $\E\left(\conddist\right)$ (we'll drop the subscripts from
now on, when the meaning is clear). In the CTR estimation problem,
for example, we can estimate $\E\left(\conddist=\baseestimate_{0}\right)$
using the average CTR of query-ad pairs with $\baseestimate$ close
to $\baseestimate_{0}$. Calibration can improve any regression method's
estimates by removing any aggregate bias. This lets us use methods
that, because of regularization, other bias-variance tradeoffs, or
model mis-specification, are efficient but biased. We can also use
calibration to turn scores, that are not estimates of $\param$ but
have information about $\param$, into estimates of $\param$ without
aggregate bias. Finally, calibration satisfies the three requirements
for the CTR estimation problem - it scales easily, it doesn't depend
on the underlying system, and it shares information by using all the
items with a similar $\baseestimate$ to estimate the correction at
that $\baseestimate$.

\begin{figure}
\includegraphics[width=1\textwidth]{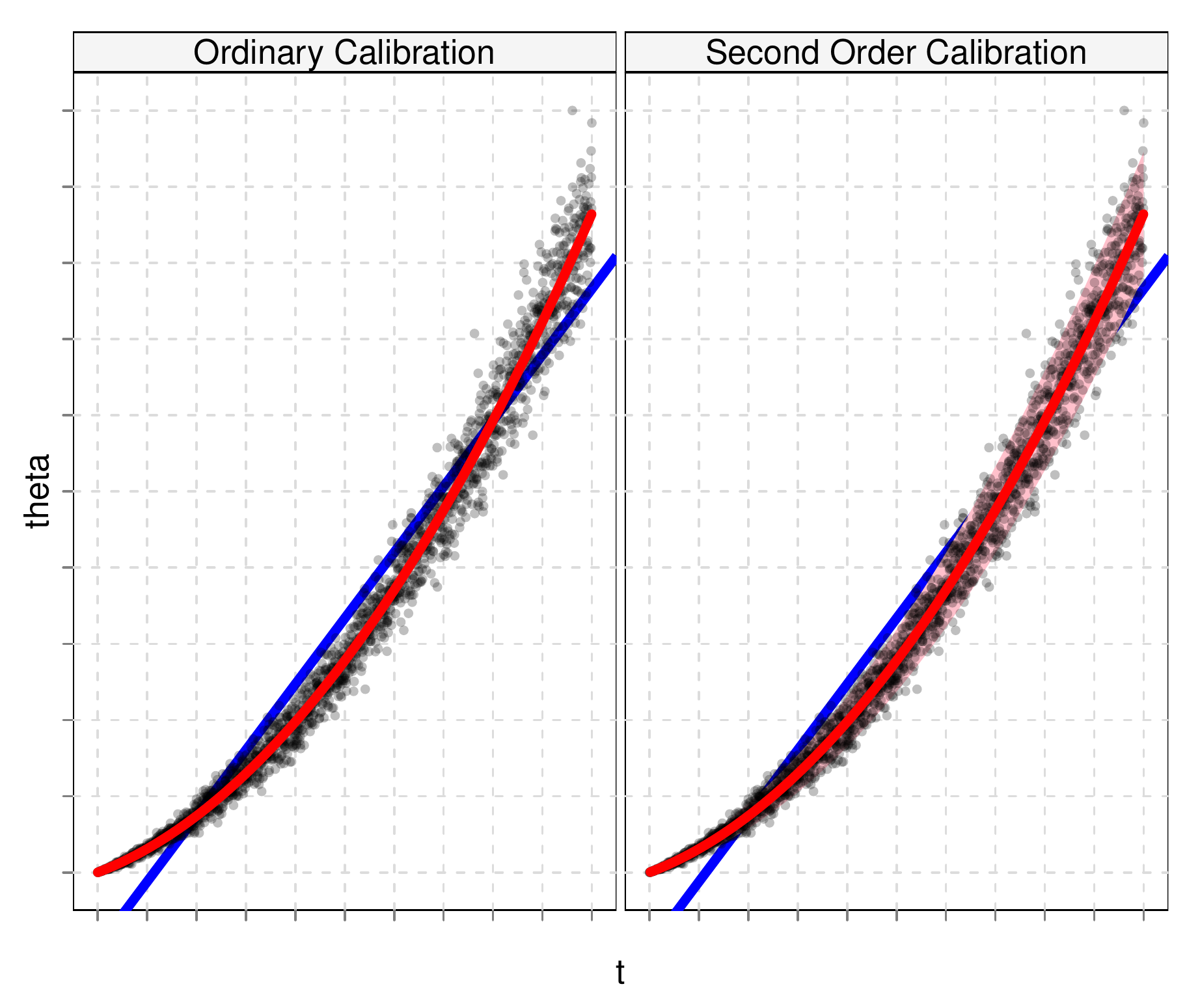}

\protect\caption{\label{fig:second-order-illustration}Ordinary and second order calibration.
Each panel plots $\protect\param$ on the $y$-axis against $\protect\baseestimate$
on the $x$-axis. Ordinary calibration, left, adjusts our estimate
from $\protect\baseestimate$ (the blue line, $x=y$) to $\protect\E\left(\protect\conddist\right)$
(the red line). Second order calibration, right, goes further. It
uses $y$ to estimate the distribution of $\protect\conddist$ (represented
by the pink color strip), and uses that distribution to estimate $\protect\Var\left(\protect\conddist\right)$,
$\protect\E\left(\protect\conddist,y\right)$ and $\protect\Var\left(\protect\conddist,y\right)$.}

\end{figure}

Calibration estimates $\E\left(\conddist\right)$. We propose estimating
the distribution of $\conddist$, and using this to approximate the
distribution of $\param\given\boldsymbol{x}$. Figure\textbf{ }\ref{fig:second-order-illustration}
illustrates the idea by plotting $\param$ vs $\baseestimate$. Calibration
adjusts our estimate, as a function of $\baseestimate$, by moving
from the $x=y$ line to the conditional mean curve $\E\left(\conddist\right)$.
The proposed method, which we call \emph{second order calibration},
goes further and estimates the distribution of $\param$ around that
curve. We don't observe the true $\param$, so we can't estimate the
distribution of $\conddist$ directly. But if many items have an estimate
close to $\baseestimate$, we can estimate the $\conddist$ distribution
by using the observed $y$s for those nearby items and employing standard
empirical Bayes techniques. Like ordinary calibration, our method
can also be used when $\baseestimate$ is a score that has information
about $\param$, rather than itself an estimate of $\param$, though
for this paper, we assume $\baseestimate$ is an estimate of $\param$.

How is this useful? Although we think approximate posteriors will
be useful in many ways, we focus on three applications: overall accuracy
estimation, improved point estimates, and fine-grained accuracy estimation.

\subsubsection*{Overall Accuracy Estimation}

Second order calibration estimates $\Var\left(\conddist\right)$,
which measures the accuracy of the estimation system (if $\baseestimate$
is a score, this measures the accuracy of the calibrated score). This
lets us separate errors due to noise, which would happen even with
perfect parameter estimates, from errors in estimation. Separating
these errors can be valuable. For example, consider the CTR estimation
problem. Because of the Poisson noise, we cannot make perfect predictions
even if we estimate $\param$ perfectly for each query-ad pair. If
we have an estimation system that predicts badly, second order calibration
can tell us whether this is because the system is inaccurate and can
be improved, or whether it is predicting as well as the noise will
allow.

\subsubsection*{Improved Point Estimates}

Second order calibration can improve point estimates through better
shrinkage. Suppose that for each item $i$, $\baseestimate_{i}$ is
trained on data independent of $y_{i}$; we can do this by dividing
our data into multiple folds, like in cross-validation, and would
need to do this anyway for ordinary calibration. Ordinary calibration
improves on $\baseestimate$ by using $\E\left(\conddist\right)$
instead. We can do even better by using $\E\left(\conddist,y\right)$.
This estimator essentially decomposes memorization and generalization.
The underlying regression method handles generalization: the distribution
of $\conddist$ reflects what we know about $\param$ based on $t$,
the underlying regression method's summary of the information in $\x$
and the other items. Second order calibration then handles memorization
by combining $\conddist$ with the item-specific information in $y$.
By estimating the distribution of $\conddist$, we can often combine
the two sources of information close to optimally. Requiring $\baseestimate_{i}$
not be trained on $y_{i}$ makes sure we don't double count the information
in $y_{i}$. In practice, this condition can be relaxed: $\baseestimate_{i}$
can be trained on $y_{i}$ as long as $y_{i}$ does not influence
$\baseestimate_{i}$ too much.

Decomposing memorization and generalization can be much better than
letting the underlying estimation system train on all the data and
handle both memorization and generalization. More interestingly, we
could design our estimation system to work with this decomposition.
For example, we could use a relatively coarse generalization model
to generate $\baseestimate$ and memorize item-specific information
using $\E\left(\conddist,y\right)$.

\subsubsection*{Fine-Grained Accuracy Estimation}

Second order calibration can estimate our accuracy for each item.
Again, suppose $\baseestimate_{i}$ is trained on data independent
of $y_{i}$ for each $i$ (or, in practice, that $y_{i}$ does not
influence $\baseestimate_{i}$ too much). We can use $\Var\left(\conddist,y\right)$
to measure the accuracy of $\E\left(\conddist,y\right)$, and as a
general measure of how much we know about each item. Second order
calibration gives us estimates of $\Var\left(\conddist,y\right)$,
which we can use to make risk-adjusted decisions and explore-exploit
tradeoffs, or to find where the underlying regression method is particularly
good or bad.

\subsubsection*{Worries and Limitations}

We might worry that there just isn't enough information in the $y$s
to get a useful estimate of the distribution of $\conddist$. If this
were true, second order calibration could never be useful. Fortunately,
this is not the case for each of the three applications above. We
prove as long as our estimate of the distribution of $y\given\baseestimate$
fits well, second order calibration will correctly estimate the true
$\Var\left(\conddist\right)$, $\E\left(\conddist\right)$ and $\Var\left(\conddist,y\right)$.
We discuss a simple diagnostic to check the fit.

Like ordinary calibration, second order calibration is intended to
be easy and useful, not comprehensive or optimal, and it shares some
of ordinary calibration's limitations. Both ordinary calibration and
second order calibration require that each $\baseestimate_{i}$ be
trained on data independent of $y_{i}$. This is easy to achieve in
principle with folds, but can be inconvenient; in practice, both methods
work well if $y_{i}$ does not influence $\baseestimate_{i}$ too
much. Both methods can also be wrong for slices of the data while
being correct on average, since they only use $\x$ through $\baseestimate$.
This is especially important for second order calibration, since we
always approximate $\param\given\x$ using $\param\given\baseestimate$.
We can guard against this by calibrating separately for important
classes of items (for example, we can calibrate CTR estimates separately
for each country), but that cannot solve the problem completely.

Second order calibration also has another important limitation, not
shared with ordinary calibration: we must have a known parametric
model for the distribution of $y\given\param$. In our CTR example,
for instance, $y\given\param$ is assumed to be $\Poisson\left(\param N\right)$,
with $N$ known. We need to know the noise mechanism to work backward
from the observed distribution of the $y$s to the $\conddist$ distribution.
It can be hard to check whether our noise model is accurate - in our
CTR example, we use predictive distributions to check the Poisson
model indirectly. Also, although we expect second order calibration
to work well for fairly general known noise, our theory only applies
when the noise is Poisson or a continuous natural exponential family
(e.g. normal with known variance).

The rest of this paper is organized as follows. We discuss related
work in Section \ref{sec:Related-Literature}. We then present our
method for second order calibration in Section \ref{sec:Method},
using CTR estimation to illustrate. In Section \ref{sub:Theoretical-Support},
we state a theoretical result justifying the three applications of
second order calibration mentioned above. Finally, in Section \ref{sec:Results-on-Real},
we present our results on the CTR estimation problem and on simulations.
Proofs are in Appendix \ref{sec:Proofs}.

\section{Related work\label{sec:Related-Literature}}

Second order calibration is closely related to existing calibration
and empirical Bayes methods.

\subsection{Calibration}

Calibration is usually used to post-process the output of good classifiers
that produce bad class probability estimates \citep{NiculescuMizil2005}.
\citet{Cohen2004} show that, in general, calibration does not reduce
classification accuracy, and makes it easier to find the right threshold
to minimize classification error. Calibration can be very effective:
\citet{Caruana2006} show that it turns boosted trees into excellent
probability estimators. The two most common methods for calibration
are Platt scaling \citep{Platt1999}, which is equivalent to logistic
regression, and isotonic regression \citep{Zadrozny2002}.

Most of the literature on calibration discusses classifiers, but regression
methods are commonly calibrated as well. Both isotonic regression
and Platt scaling generalize straightforwardly to calibrating regression
methods. \citet{Amini2009} use random effect methods to calibrate
and estimate the accuracy of regression methods.

\subsection{Empirical Bayes}

Empirical Bayes methods use Bayesian inference to solve problems,
but estimate priors instead of using subjective or reference priors.
The key idea is that if we have many independent draws from a model
with unknown prior, we can use the data to estimate the prior, or
a quantity of interest that depends on the prior.

For example, suppose that $\mu$ comes from an unknown prior $G$,
our data $z$ is $\mathcal{N}\left(\mu,1\right)$, and we observe
many $z$s from this model. We want to say something about the unobserved
$\mu_{i}$ corresponding to each $z_{i}$. As \citet{Robbins1954}
showed, we can use the $z$s to estimate the prior, then estimate
$\mu_{i}$ using $\E_{\hat{G}}\left(\mu_{i}\given z_{i}\right)$,
where $\E_{\hat{G}}$ is the expectation in our model under the estimated
prior $\hat{G}$. This estimate of $\mu_{i}$ combines global information
from all the $z$s (via $\hat{G}$) with the specific information
in $z_{i}$.

Emprical Bayes methods work when the quantity we're interested in
can be expressed in terms of the marginal density of the data. Such
an expression tells us we can estimate the quantity, since we can
estimate the marginal density using our data. In our normal example,
Tweedie's formula \citep{Robbins1954} shows that if $f_{G}$ is the
marginal density of the $z$s,
\[
\E_{G}\left(\mu\given z\right)=z-\frac{f_{G}'\left(z\right)}{f_{G}\left(z\right)}
\]
for any prior $G$ (Robbins actually estimated this quantity directly
instead of using an estimate of the prior). Since we observe many
$z$s, we can estimate the marginal and thus estimate $\E_{G}\left(\mu\given z\right)$
without knowing the prior in advance.

Empirical Bayes methods need a large amount of data to shine. Big
data sets have made them increasingly useful; \citet{Efron2010} gives
an introduction and review. They have enjoyed particular success recently
in signal processing (e.g. \citep{Johnstone2004}) and multiple testing
(though false discovery rates, e.g. \citep{Efron2001}).

\subsection{Why not bootstrap?}

At first glance, the bootstrap seems like a natural way to get approximate
posteriors for any regression method. Before we present second order
calibration, it is worth understanding why the bootstrap doesn't work
for this problem. The bootstrap is often too slow for large data sets,
since it requires training the regression method many times. In the
CTR estimation problem, for example, bootstrapping would require training
a massive, resource-intensive machine learning system tens or even
hundreds of times. This is impractical and expensive. Post-processing
methods like ordinary and second order calibration are much easier
to use.

More importantly, though, the bootstrap estimates the distribution
of $\baseestimate$, not of $\conddist$, and these distributions
can be very different, particularly for the biased estimators often
used on large data sets. Consider the trivial estimator $t=0$. The
bootstrap would correctly find that $t$ is always $0$, but this
tells us nothing about the distribution of $\conddist$.

\section{The Proposed Method\label{sec:Method}}

We now state the second order calibration problem more precisely.
We are given responses $y_{i}$ for items $i=1,\ldots,I$. Each item
has a parameter $\param_{i}$ that controls the response in a known
way: $y_{i}\given\param_{i}\sim f_{\param_{i}}$, where $f_{\param}$
is a given parametric family. The family $f_{\theta}$ can depend
on known offsets that are different for each $i$; we suppress this
dependence in our notation. We assume that $f_{\param}$ is either
Poisson ($y\given\param\sim\Poisson\left(N\param\right)$) or a continuous
exponential family with natural parameter $\param$ (for example,
$y\given\param\sim\mathcal{N}\left(\param,\sigma^{2}\right)$ with
known $\sigma$). For each item $i$, a machine learning system gives
us an estimate $\baseestimate_{i}$ that is not trained on $y_{i}$.
Our goal is to estimate the posterior distribution $\conddist$, which
we denote by $\prior$, for each $t$.

We will use $\prior$ to approximate the full posterior distribution
$\param\given\x$. The quality of this approximation will depend on
the underlying machine learning method and the data set. In this paper,
we will not try to quantify the approximation error. Our goal is to
estimate functionals of the true $\prior$, which average the true
$\param\given\x$ distributions, in the same way that ordinary calibration
estimates $\E\left(\conddist\right)$, not $\E\left(\param\given\x\right)$,
and is content to be correct on average.

\subsection{In a Nutshell}

The proposed method has five steps:
\begin{enumerate}
\item Bin the items by $\baseestimate$, so that $\baseestimate$ is approximately
constant in each bin.
\item Estimate $\prior$ separately for each bin. Assume $\prior$ is constant
in the bin, so the observations in the bins are drawn from the model
\begin{eqnarray}
\param\given t & \sim & G\nonumber \\
y\given\param,t & \sim & f_{\param},\label{eq:in-bin-model}
\end{eqnarray}
where $G\equiv G_{t}$ is the common value of $\prior$ for items
in the bin. Choose a parametrization for $G$, and estimate $G$ by
maximum marginal likelihood. Do this in parallel for all the bins
to get an estimate $\priorestimate$ for each bin.
\item Collect the $\priorestimate$. If necessary, adjust them so that $\E_{\hat{G}}\left(\param\given\baseestimate\right)$
and $\Var_{\hat{G}}\left(\param\given\baseestimate\right)$ are smooth
functions of $\baseestimate$. Here, $\E_{\hat{G}}\left(\cdot\given t\right)$
and $\Var_{\hat{G}}\left(\cdot\given t\right)$ denote the expectation
and variance in Model \ref{eq:in-bin-model}, above, with prior $\priorestimate$.
\item Check the fit. When plugged into Model \ref{eq:in-bin-model}, the
adjusted $\priorestimate$ should lead to marginal distributions of
$y\given\baseestimate$ that fit the data.
\item Calculate the overall calibration curve $\hat{\E}\left(\conddist\right)=\E_{\hat{G}}\left(\conddist\right)$,
overall accuracy curve $\hat{\Var}\left(\conddist\right)=\Var_{\hat{G}}\left(\conddist\right)$,
updated estimates $\hat{\E}\left(\conddist,y\right)=\E_{\hat{G}}\left(\conddist,y\right)$
and fine-grained accuracy estimates $\hat{\Var}\left(\conddist,y\right)=\Var_{\hat{G}}\left(\conddist,y\right)$.
\end{enumerate}
If the items naturally fall into coarse categories, we can follow
these steps separately for each category. For example, in the CTR
estimation problem, we can treat the query-ad pairs for each country
separately.

In the rest of this section, we discuss each step in more detail,
illustrating with the CTR estimation problem.

\subsection{Step 1: Binning by $\protect\baseestimate$}

Binning the items by $\baseestimate$ is straightforward - simply
divide the range of $\baseestimate$ into $B$ bins. The only questions
are how to choose $B$, and how to set the bin boundaries. Using quantiles
of the distribution of $t$ for the bin boundaries seems to work well.
This gives bins with the same number of items.

Choosing $B$ is a bias-variance tradeoff: each bin has to be small
enough so that $\baseestimate$ is approximately constant in each
bin, but big enough so that we have enough data in each bin to estimate
$\prior$. Because we later smooth our estimates of $\prior$ across
bins, the choice of bin width is not crucial, as long as it is in
a reasonable range. For CTR estimation, we tried different choices
of $B$ and judged them by how wide the bins were and how stable the
fitted $\priorestimate$ were. In the end, we found that a range of
$B$s all gave reasonably well-behaved $\priorestimate$, and, for
maximum parallelism, chose the largest reasonable $B$.

\subsection{Step 2: Fitting $\protect\prior$ for each bin}

We now work within a single bin. Within this bin, $\baseestimate$
is approximately constant, so all the $\prior$ are all approximately
the same distribution, $G$. This means that the data in the bin approximately
come from the model
\begin{eqnarray*}
\param\given t & \sim & G\\
y\given\param,t & \sim & f_{\param}.
\end{eqnarray*}
We estimate $G$ by giving it a convenient parametrization and estimating
the parameters by maximum marginal likelihood. That is, we maximize
the marginal log-likelihood 
\[
\sum\log f_{G}\left(y\right)
\]
where the sum is taken over the items in the bin, and 
\[
f_{G}\left(y\right)=\int f_{\param}\left(y\right)dG\left(\param\right)
\]
is the marginal density of $y$ that corresponds to $G$. The distribution
$f_{G}=f_{G_{t}}$ depends on $\baseestimate$ (through the bin) and
on any known offsets in $f_{\param}$, but our notation suppresses
this.

In the CTR estimation problem, for example, we model $G$ using a
Gamma distribution. This makes $f_{G}$ negative binomial, with mean
and dispersion that depend on $N$ and the shape and scale of $G$.
We fit $G$'s shape and scale by finding the values that maximize
the negative binomial likelihood.

How should we model $G$? The theoretical results in Section \ref{sub:Theoretical-Support}
show that the details of the choice aren't too important, at least
for the applications in this paper. What matters is that $\hat{G}$
leads to a marginal distribution that fits the observed data; that
guarantees the final results will be correct on average. This means
we should choose the simplest, most convenient model for $G$ that
fits the data.

We recommend first trying to model $G$ as a conjugate prior. If that
proves too restrictive, we recommend modeling $G$ as a mixture of
conjugate priors, using the simplest model necessary to fit the data
(use the fewest or otherwise most constrained mixture components).
Conjugate prior mixtures and similar nonparametric maximum likelihood
methods often perform well in empirical Bayes problems \citep{Kiefer1956,Muralidharan2010,Jiang2009}.
They are flexible enough to fit any distribution, with enough components,
but can still be manipulated using conjugacy formulas. For the CTR
estimation problem, we tried two models for $G$ - a simple Gamma
distribution, and a mixture of Gammas. For the latter, we fixed the
mixture components and fit the weights using the standard EM algorithm
for mixtures. We found that a single Gamma distribution fit our data
well (Subsection \ref{sub:Real-CTR-data} examines the fit).

Our method scales to large data sets easily because we find $\priorestimate$
separately for each bin, with no communication between bins. This
lets us find $\priorestimate$ for all the bins in parallel. Since
we only need $y$ (and $N$ for CTR estimation) for items in a bin
to find $\priorestimate$, each bin can usually be handled by one
machine and that makes parallelization especially easy.

\subsection{Step 3: Adjusting $\protect\priorestimate$ if necessary}

The disadvantage of fitting in each bin separately is that $\priorestimate$
may not vary nicely with $\baseestimate$. For example, $\E_{\hat{G}}\left(\conddist\right)$
and $\Var_{\hat{G}}\left(\conddist\right)$ may not be smooth functions
of $\baseestimate$. We may also want $\E_{\hat{G}}\left(\conddist\right)$
to be a monotone function of $\baseestimate$. Sometimes this isn't
a problem - if we have enough data in each bin, the $\priorestimate$
can vary nicely enough with $\baseestimate$ even though we haven't
constrained them to do so.

If not, though, we can fix the problem by smoothing or monotone regression.
For CTR estimation, we used smoothing splines to get smoothed versions
of $\E_{\hat{G}}\left(\conddist\right)$ and $\Var_{\hat{G}}\left(\conddist\right)$,
as functions of $\baseestimate$, then adjusted the $\priorestimate$
so their means and variances matched the smoothed $\E_{\hat{G}}\left(\conddist\right)$
and $\Var_{\hat{G}}\left(\conddist\right)$. To make sure $\priorestimate$
stayed positive, we shifted and scaled the distributions of $\log\param$
instead of $\param$.

\subsection{Step 4: Checking the fit}

Our theoretical results show that we must fit the marginal distribution
of the data well to get good results. This means we need to check
the marginal fit before we use the $\priorestimate$. Let $\hat{f}_{\baseestimate}$
be the marginal distribution of $y\given\baseestimate$ in the within-bin
model ($\param\given t\sim\priorestimate$, $y\given t,\param\sim f_{\param}$).
The distribution $\hat{f}_{\baseestimate}$ depends on any known offsets
in $f_{\param}$, but our notation suppresses this. We need to check
that the $\hat{f}_{\baseestimate}$s fit the observed $y$s.

There are many ways to assess the fit of a collection of distributions
(see \citep{Gneiting2007} for a discussion of different criteria,
in the context of predictive distributions). We suggest using the
standard probability integral transform, randomized to account for
the discreteness of $y$. Let
\[
p=\P_{\hat{f}_{\baseestimate}}\left(Y\leq y\right)-u\P_{\hat{f}_{\baseestimate}}\left(Y=y\right)
\]
where $u\sim\Uniform\left(0,1\right)$ is independent of $y$, and
$\P_{\hat{f}_{\baseestimate}}$ is probability under $Y\sim\hat{f}_{\baseestimate}$.
Each $p$ is $\Uniform\left(0,1\right)$ if and only if $\hat{f}_{\baseestimate}$
is the true distribution of $y\given\baseestimate$, and the more
non-uniform $p$ is, the more $\hat{f}_{\baseestimate}$ differs from
the true distribution of $y\given\baseestimate$ \citep{Muralidharan2012}.

We check the fit of the $\priorestimate$ by looking at the observed
distribution of the $p$s in each bin. If the $p$s are non-uniform
in a bin, then $\priorestimate$ does not fit the data in that bin
well. When this happens, the shape of the $p$ histogram often suggests
a solution. For example, a U-shaped histogram says that $\priorestimate$
is too light-tailed, since we see more large and small $p$s than
$\hat{G}_{t}$ predicts. If the $p$s are uniform in each bin, our
fits are at least correct on average, so the theory says we can expect
reasonable results.

\subsection{Step 5: Calculating useful quantities}

Armed with $\priorestimate$, we can now compute the overall calibration
curve $\hat{\E}\left(\conddist\right)$, overall accuracy curve $\hat{\Var}\left(\conddist\right)$,
updated estimates $\hat{\E}\left(\conddist,y\right)$ and fine-grained
accuracy estimates $\hat{\Var}\left(\conddist,y\right)$. These can
all be computed quickly in closed form if $\priorestimate$ is a conjugate
prior or conjugate mixture. Each quantity only involves $\priorestimate$,
$y$ and a known offset like $N$ for one item, so we can treat the
items in parallel.

\section{Theoretical Support\label{sub:Theoretical-Support}}

We now present a simple theoretical result that says second order
calibration will give us good estimates of $\E\left(\conddist\right)$,
$\Var\left(\conddist\right)$, $\E\left(\conddist,y\right)$ and $\Var\left(\conddist,y\right)$
as long as we fit the marginal distribution of $y\given\baseestimate$.

The result tries to address two worries. First, we might worry that
the $y$s just don't have enough information to estimate the quantities
we are interested in. For example, suppose $y\sim\mathcal{N}\left(\param,1\right)$,
and we tried to use the $y$s to estimate $\P\left(\param=0\given\baseestimate\right)$.
This is essentially impossible: we will never have enough data to
choose between the two distributions $\conddist=0$ and $\conddist=\varepsilon$
for small enough $\varepsilon$, since they produce very similar marginal
distributions of $y\given\baseestimate$, but the two distributions
lead to very different estimates of $\P\left(\param=0\given\baseestimate\right)$.
We need to show that second order calibration does not fall into the
same trap.

Second, we might worry that the exact model we use for $\prior$ will
strongly influence our results. If so, this would be a serious problem,
since we have no principled way to choose between two models that
fit the data equally well.

It turns out that neither of these worries is a problem, at least
when $f_{\param}$ is Poisson or a continuous natural exponential
family. The $y$'s have enough information to estimate $\E\left(\conddist\right)$,
$\Var\left(\conddist\right)$, $\E\left(\conddist,y\right)$ and $\Var\left(\conddist,y\right)$,
and any model for $\prior$ that fits the data well will give similar
estimates for these four quantities.

This happens because each quantity can be written in terms of the
marginal distribution of $y\given\baseestimate$. We can learn the
marginal using data, and models with the same marginal give the same
estimates. For example, \citet{Robbins1954} shows that if $y\sim\Poisson\left(\param N\right)$,
\begin{eqnarray*}
\E\left(\conddist,y\right) & = & \frac{y+1}{N}\frac{f_{G}\left(y+1\right)}{f_{G}\left(y\right)}\\
\Var\left(\conddist,y\right) & = & \frac{\left(y+2\right)\left(y+1\right)}{N^{2}}\frac{f_{G}\left(y+2\right)}{f_{G}\left(y\right)}-\left(\frac{y+1}{N}\frac{f_{G}\left(y+1\right)}{f_{G}\left(y\right)}\right)^{2},
\end{eqnarray*}
where $G=\prior$, and $f_{G}$ is the marginal distribution of $y\given\baseestimate$
in Model \ref{eq:in-bin-model} (for simplicity, we drop the $t$
in the subscript instead of writing $f_{\prior}$). We can express
$\E\left(\conddist\right)$ and $\Var\left(\conddist\right)$ in terms
of the marginal by taking the expectations of $\E\left(\conddist,y\right)$
and $\Var\left(\conddist,y\right)$ over $y$ and using the conditional
variance identity. Similar formulas for continuous natural exponential
families are in the Appendix.

There is one caveat that is theoretically important, though not practically.
The formulas all involve dividing by $f_{G}\left(y\right)$, so they
can behave badly if $f_{G}\left(y\right)$ is close to zero. If our
estimated $\priorestimate$ gives a marginal with light tails, our
estimates can behave badly. To guard against this problem, we can
regularize our estimates by dividing by $ $$\max\left(f_{\hat{G}}\left(y\right),\rho\right)$
instead of by $f_{\hat{G}}\left(y\right)$, where $\rho$ is a tuning
parameter \citep{Zhang1997}. This is slightly unnatural, but the
resulting regularized estimators actually have some nice properties
that we discuss in the Appendix. We find regularization unnecessary
in practice - the $\priorestimate$ that fit our data usually aren't
light-tailed. Should we need to regularize, we can choose $\rho$
to maximize the predictive accuracy of the regularized estimator on
a test set.

Theorem \ref{thm:error-bound} makes this marginal distribution argument
more precise. Building on regret bounds in the empirical Bayes literature
\citep{Jiang2009,Muralidharan2011a}, it bounds the error in our estimates
of $\E\left(\conddist,y\right)$ and $\Var\left(\conddist,y\right)$
in terms of our error in estimating the marginal density and its derivatives,
plus a regularization term that vanishes as $\rho\rightarrow0$. Because
we can use these to get $\E\left(\conddist\right)$ and $\Var\left(\conddist\right)$,
the theorem implies that the error in our estimates of the latter
two can also be bounded in terms of our error in estimating the marginal.

Bounds like the ones in the theorem can be used to find rates of convergence
for empirical Bayes estimators, but we think it serves better as motivation
and a sanity check for second order calibration than as a technical
tool. Its bounds are for the regularized estimates; the regularized
and unregularized estimates are usually very close, but for completeness,
we give bounds for the unregularized estimates in the Appendix. The
Appendix also has error bounds for estimates of higher cumulants,
like the skewness and kurtosis. 

We measure error using the $L^{2}$ norm, weighted by $f_{G}$: $\left\Vert h\right\Vert =\left(\int h\left(y\right)^{2}f_{G}\left(y\right)dy\right)^{\frac{1}{2}}$.
For the Poisson, the theorem gives bounds for $\E\left(\theta\given t,y\right)$
and $\E\left(\theta^{2}\given t,y\right)$, which is clearly equivalent
to bounding the errors for $\E\left(\theta\given t,y\right)$ and
$\Var\left(\theta\given t,y\right)$.
\begin{thm}
\label{thm:error-bound}

If $f_{\theta}$ is $\Poisson\left(N\theta\right)$, the regularized
posterior moment estimators have error at most\textup{ 
\begin{eqnarray*}
\left\Vert \hat{\E}_{\rho}\left(\conddist,y\right)-\E\left(\conddist,y\right)\right\Vert  & \leq & \frac{C_{G,\rho}}{N}\left(\left\Vert \left(f_{\hat{G}}-f_{G}\right)^{2}\right\Vert ^{\frac{1}{2}}+\left\Vert \left(f_{\hat{G}}\left(y+1\right)-f_{G}\left(y+1\right)\right)^{2}\right\Vert ^{\frac{1}{2}}\right)+\frac{D_{G,\rho}}{N}\\
\left\Vert \hat{\E}_{\rho}\left(\theta^{2}\given\baseestimate,y\right)-\E\left(\theta^{2}\given\baseestimate,y\right)\right\Vert  & \leq & \frac{C_{G,\rho}}{N^{2}}\left(\left\Vert \left(f_{\hat{G}}-f_{G}\right)^{2}\right\Vert ^{\frac{1}{2}}+\left\Vert \left(f_{\hat{G}}\left(y+2\right)-f_{G}\left(y+2\right)\right)^{2}\right\Vert ^{\frac{1}{2}}\right)+\frac{D_{G,\rho}}{N^{2}}
\end{eqnarray*}
}where $C_{G,\rho}$, $D_{G,\rho}$ are constants that only depend
on $G$, $\rho$ (not the same from line to line), and $\lim_{\rho\rightarrow0}D_{G,\rho}=0$.

If $f_{\theta}$ is a continuous natural exponential family, the regularized
posterior moment estimators have error at mos\textup{t}
\begin{eqnarray*}
\left\Vert \hat{\E}_{\rho}\left(\param\given\baseestimate,y\right)-\E\left(\param\given\baseestimate,y\right)\right\Vert  & \leq & C_{G,\hat{G},\rho}\left\Vert f_{G}'-f_{\hat{G}}'\right\Vert +H_{G,\rho}\left\Vert \left(f_{G}-f_{\hat{G}}\right)^{2}\right\Vert ^{\frac{1}{2}}+D_{G,\rho}
\end{eqnarray*}
\begin{eqnarray*}
\left\Vert \hat{\Var}_{\rho}\left(\param\given\baseestimate,y\right)-\Var\left(\param\given\baseestimate,y\right)\right\Vert  & \leq & C_{G,\hat{G},\rho}\left\Vert \left(\left(f_{G}'-f_{\hat{G}}'\right)^{2}+\left(f_{G}''-f_{\hat{G}}''\right)^{2}\right)^{\frac{1}{2}}\right\Vert \\
 &  & +H_{G,\rho}\left\Vert \left(f_{G}-f_{\hat{G}}\right)^{2}\right\Vert ^{\frac{1}{2}}+D_{G,\rho}\\
\end{eqnarray*}
where $C_{G,\hat{G},\rho}$, $D_{G,\rho}$, $H_{G,\rho}$ are constants
that depend on $G$, $\rho$ (not the same from line to line) and
$\lim_{\rho\rightarrow0}D_{G,\rho}=0$. $C_{G,\hat{G},\rho}$ also
depends on $\hat{G}$ through $\left\Vert f_{\hat{G}}\right\Vert _{\infty},\left\Vert f_{\hat{G}}'\right\Vert _{\infty},\left\Vert f_{\hat{G}}''\right\Vert _{\infty}$.
\end{thm}

\section{Results on Real Data and Simulations\label{sec:Results-on-Real}}

We now show that second order calibration performs well on real and
simulated CTR data, and on a simulated normal data set. For each data
set, we show the fitted $\priorestimate$ and the $p$ histogram that
checks for model fit. We then show that second order calibration can
estimate $\E\left(\conddist\right)$, $\Var\left(\conddist\right)$,
$\E\left(\conddist,y\right)$ and $\Var\left(\conddist,y\right)$
well, and that the second order calibrated estimates $\E\left(\conddist,y\right)$
often significantly improve on the original estimates $\baseestimate$
and the first order calibrated estimates $\E\left(\conddist\right)$.
Note that because the bootstrap was computationally impractical and
doesn't estimate what we are interested in, we did not include it
as a baseline.

\subsection{Real CTR data\label{sub:Real-CTR-data}}

We first illustrate our method with real CTR data, with clicks $y_{i}$
and impressions $N_{i}$ for more than 20 billion query-ad pairs.
For each query ad-pair, we also have $\baseestimate_{i}$, an estimate
of the CTR generated by a black-box regression method. The CTR estimates
are actually generated for each impression, and we sum them to get
an overall $\baseestimate_{i}$ for each query-ad pair. This induces
a slight dependence between $\baseestimate_{i}$ and $y_{i}$ - although
the predictions and response are independent at the impression level,
the prediction for an impression may depend on previous responses,
and this creates dependence at the query-ad level. This dependence
is small for most query-ad pairs, so we ignore it.

\begin{figure}
\includegraphics[width=1\textwidth]{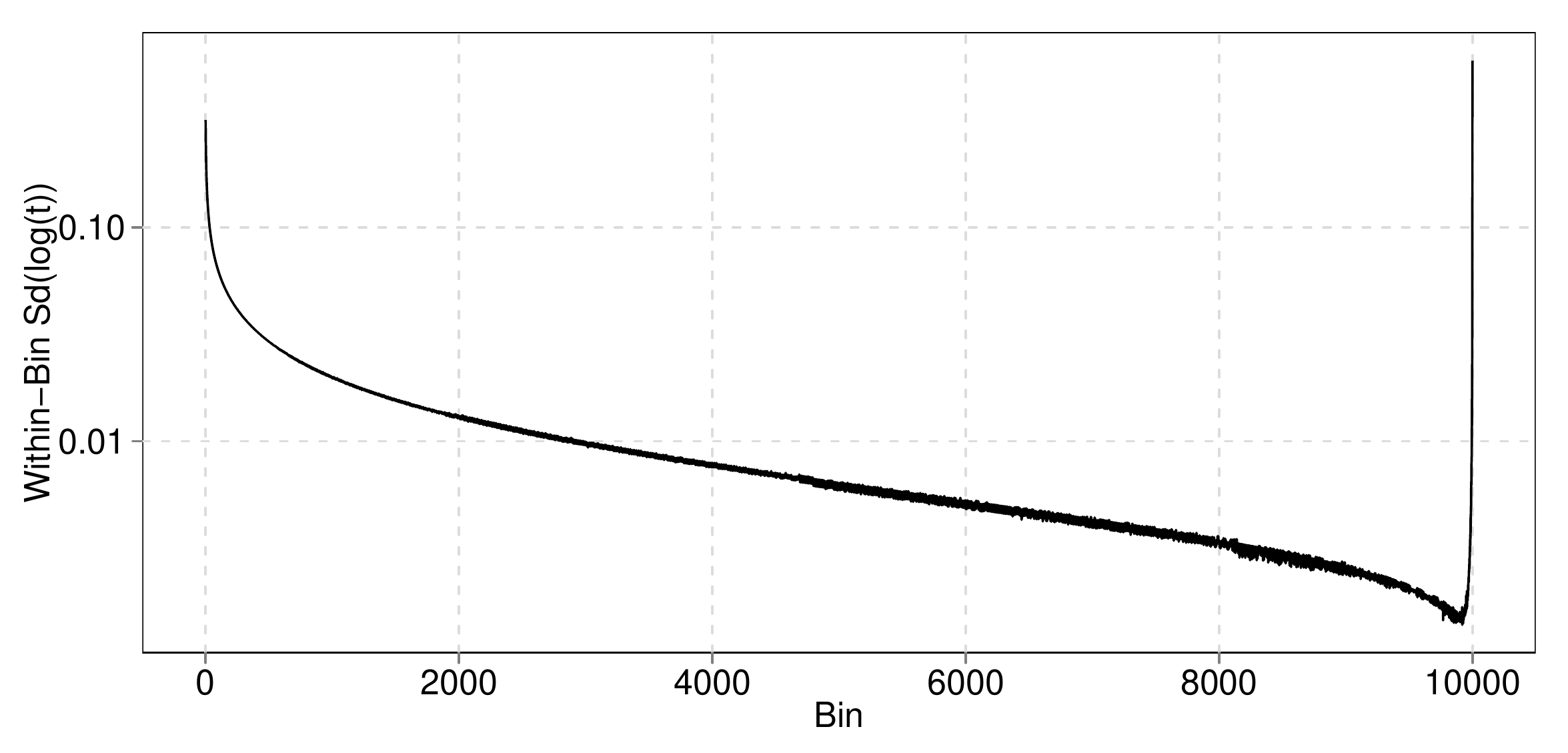}

\protect\caption{Within-bin range of $t$, measured by $\mbox{Sd}\left(\log t\right)$
in each bin, plotted on the log scale. \label{fig:Within-bin-range-of}}

\end{figure}

\begin{figure}
\includegraphics[width=1\textwidth]{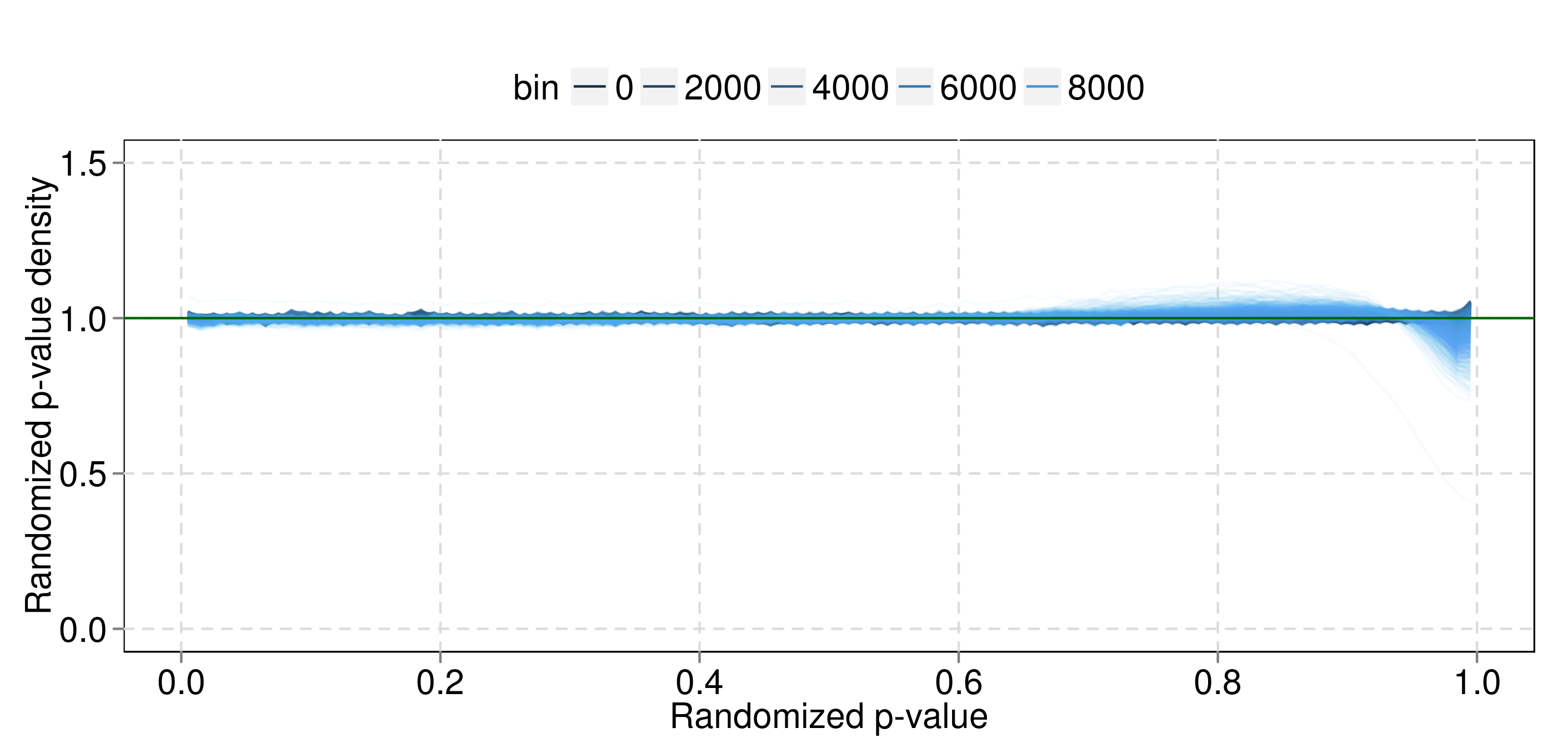}

\protect\caption{$p$-histogram densities for the marginal distribution within each
bin. The densities are mostly flat, indicating that the $\protect\priorestimate$
fit well.\label{fig:marginal-pval}}
\end{figure}

\begin{figure}
\includegraphics[width=1\textwidth]{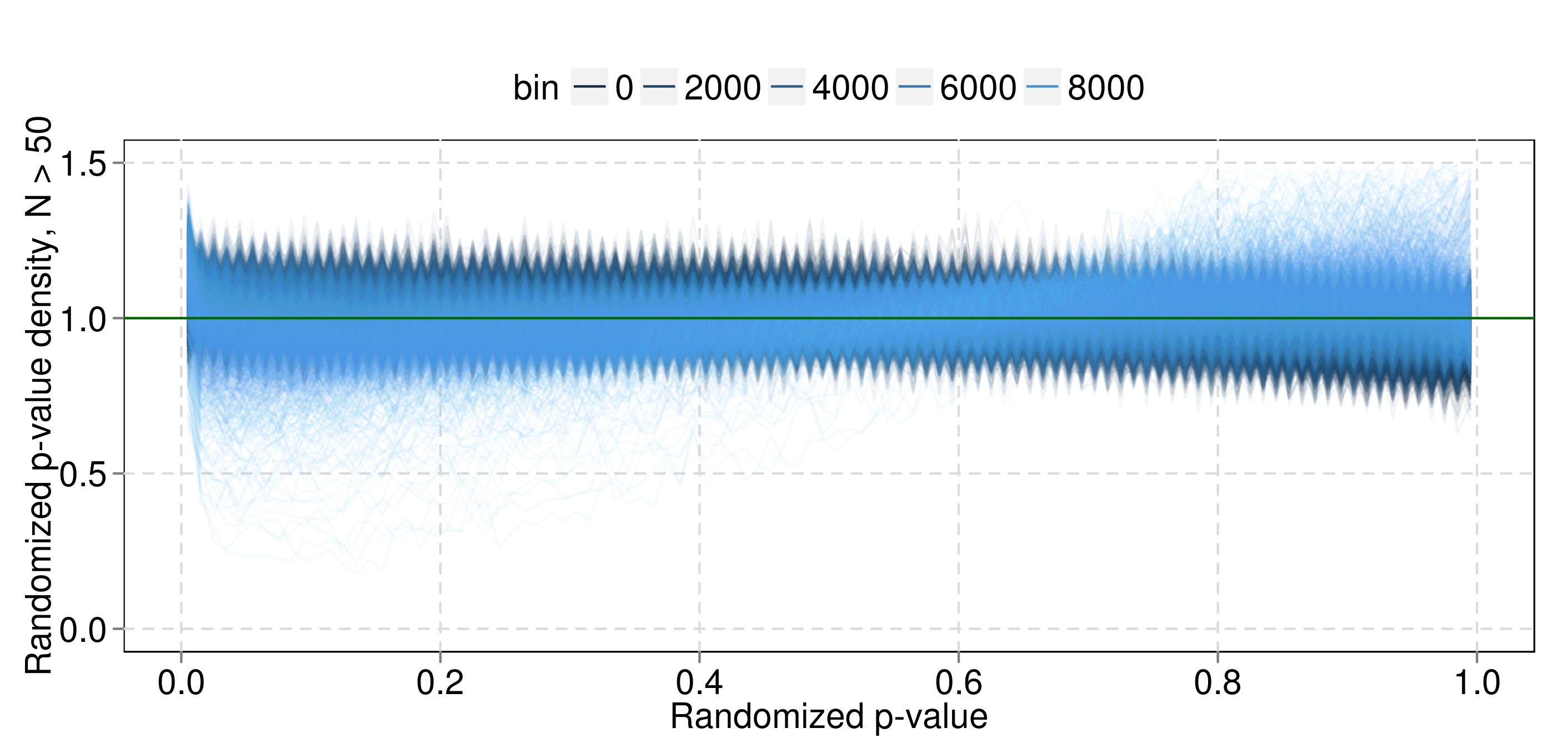}

\protect\caption{$p$-histogram densities for the marginal distribution within each
bin, restricted to query-ad pairs with $N\geq50$. The densities are
mostly flat, indicating that the $\protect\priorestimate$ fit well.
The band is wider than the band in Figure \ref{fig:marginal-pval}
because we have fewer query-ad pairs, and so noisier histograms. The
spikes are artifactual: they appear because we plot the histograms
at discrete points, so we see the full range of the curves at those
points and not in between.\label{fig:marginal-pval-bign}}
\end{figure}

We divided the query-ad pairs into 10,000 bins, each with an equal
number of clicks. Within each bin, $\baseestimate$ was approximately
constant: $\log\left(t\right)$ usually had a standard deviation of
less than $0.01$, but the left and rightmost bins are much wider
(Figure \ref{fig:Within-bin-range-of}). We tried two models for $\priorestimate$
- a Gamma distribution, and a mixture of $100$ Gamma distributions
with fixed gamma parameters (chosen to have equispaced mean and equal
variance on the log scale). Figure \ref{fig:marginal-pval} shows
the $p$ histograms for the single Gamma model. The $p$s are mostly
uniform, indicating that the model fits well; the densities are slightly
skewed right, indicating that our $\priorestimate$ don't have quite
enough mass on the right tail. The fit is also poor in the very rightmost
bins.\textbf{ }The $p$ histograms have little power when $N$ is
small, since any sensible $\priorestimate$ will give a marginal density
of $y$ concentrated at $0$. To make sure we aren't just seeing this
zero-effect, we also looked at the $p$ histogram for query-ad pairs
with $N\geq50$, where we have more power to detect if our model fits
badly (Figure \ref{fig:marginal-pval-bign}). These looked similar,
indicating that our model actually fits the data well for most bins.
Based on this, we used a single Gamma model for the rest of our analysis.

\begin{figure}
\includegraphics[width=1\textwidth]{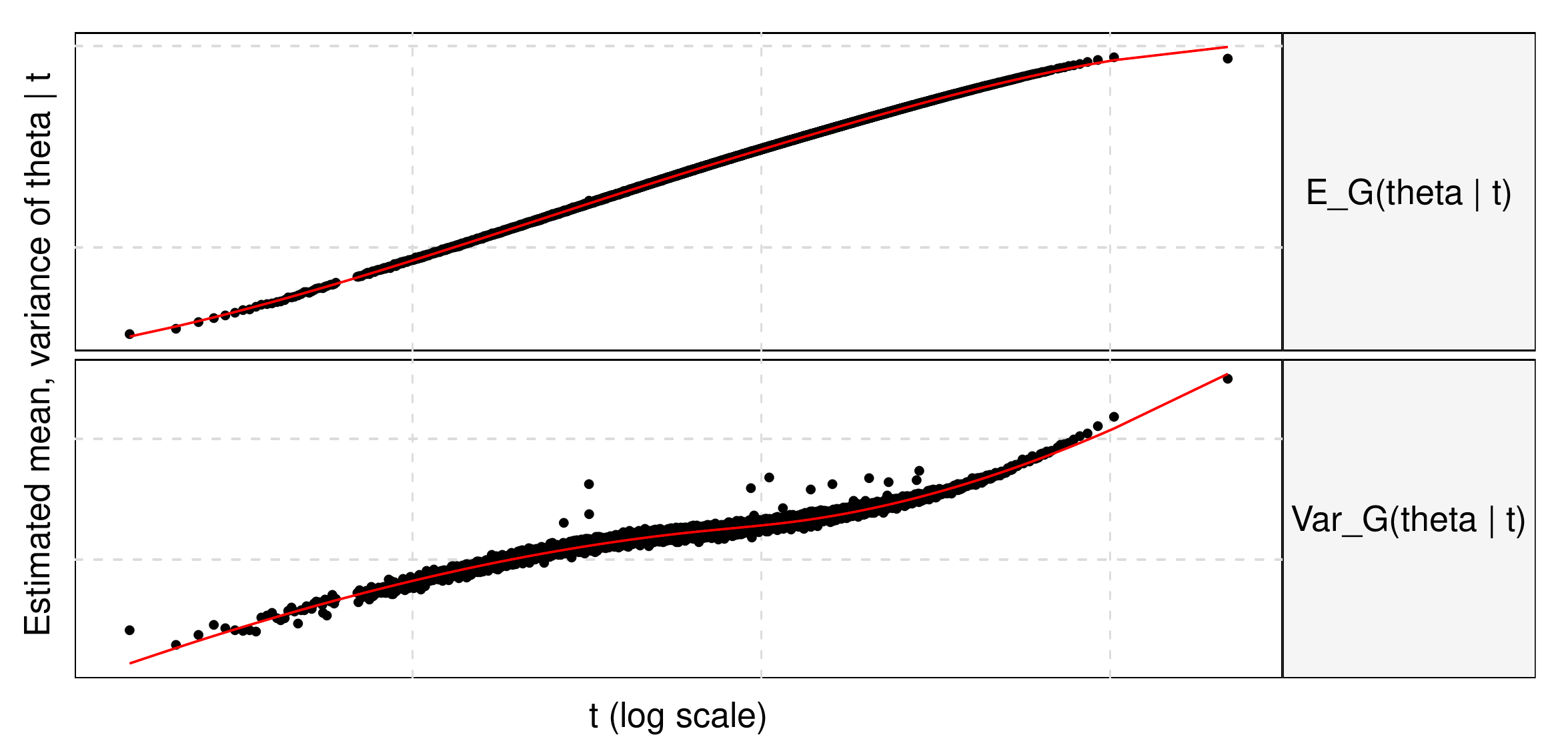}

\protect\caption{Unsmoothed (dots) and smoothed (red lines) $\protect\E_{\hat{G}}\left(\protect\conddist\right)$
and $\protect\Var_{\hat{G}}\left(\protect\conddist\right)$ (top and
bottom, respectively). $\protect\E_{\hat{G}}\left(\protect\conddist\right)$
is pretty smooth, but $\protect\Var_{\hat{G}}\left(\protect\conddist\right)$
needs smoothing.\label{fig:Unsmoothed-and-smooth}}
\end{figure}

\begin{figure}
\includegraphics[width=1\textwidth]{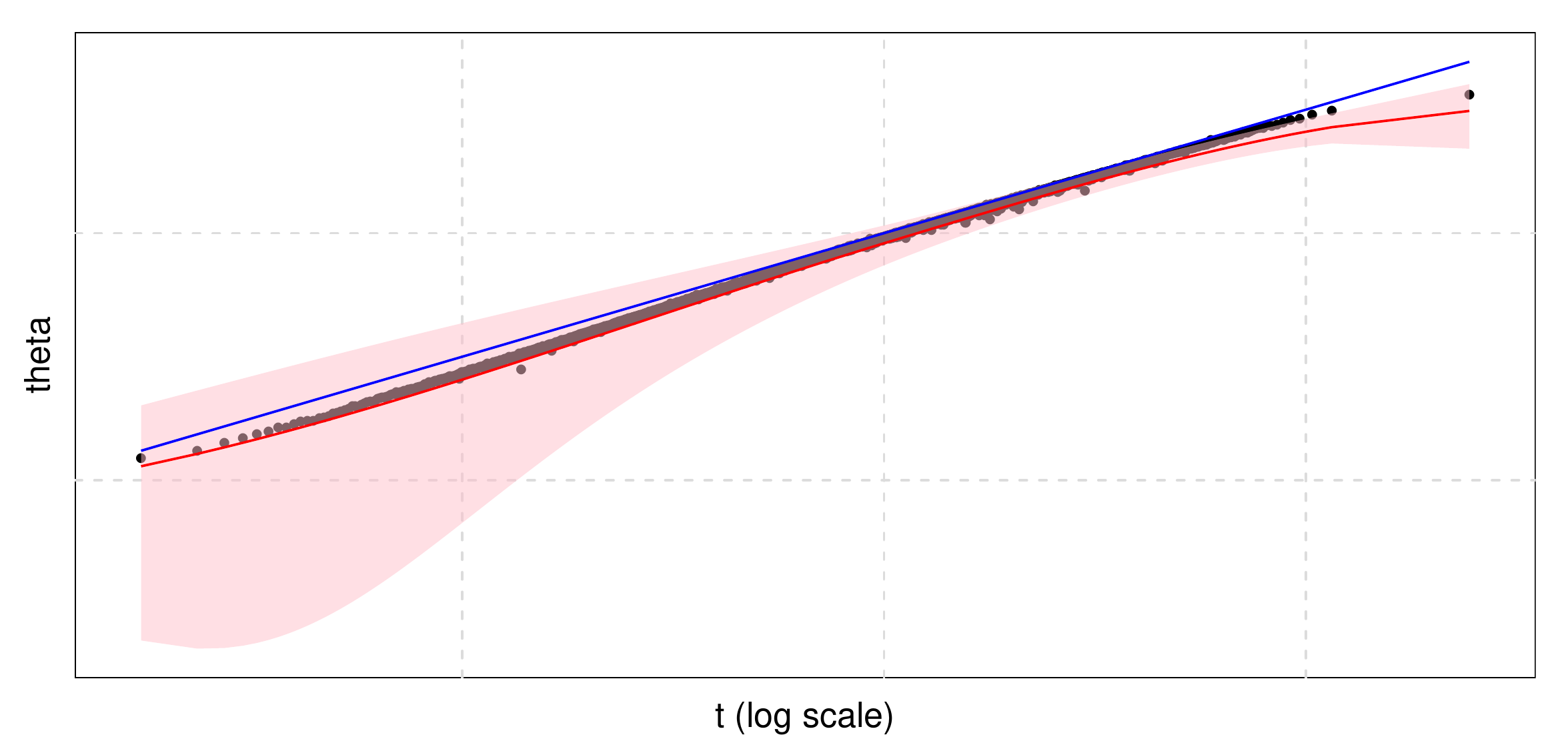}

\protect\caption{Adjusted $\protect\priorestimate$. The black dots are an overall
estimate of the mean $\protect\param$ in each bin ($\sum y/\sum N$).
The blue line is $\protect\param=\protect\baseestimate$. The red
line is the fitted $\protect\E_{\hat{G}}\left(\protect\conddist\right)$,
and the pink shaded area shows the $10$th to $95$th percentile of
$\protect\priorestimate$ at each $t$. The red curve is a little
below the black dots because the smoothing pulls it down (Figure \ref{fig:Unsmoothed-and-smooth};
the rightmost point has disproportionate leverage) , and because it
is estimating $\protect\E\left(\protect\param\right)=\protect\E\left(y/N\right)$,
which can be lower than $\sum y/\sum N=\sum\left(\frac{N}{\sum N}\right)\frac{y}{N}$
if bigger $N$s are associated with bigger $\protect\param$s. The
likelihood improvement in Figure \ref{fig:LL-Percentage-improvement}
suggests that this is a better shrinkage target.\label{fig:final-fitted-prior}}
\end{figure}

\begin{figure}
\includegraphics[width=1\textwidth]{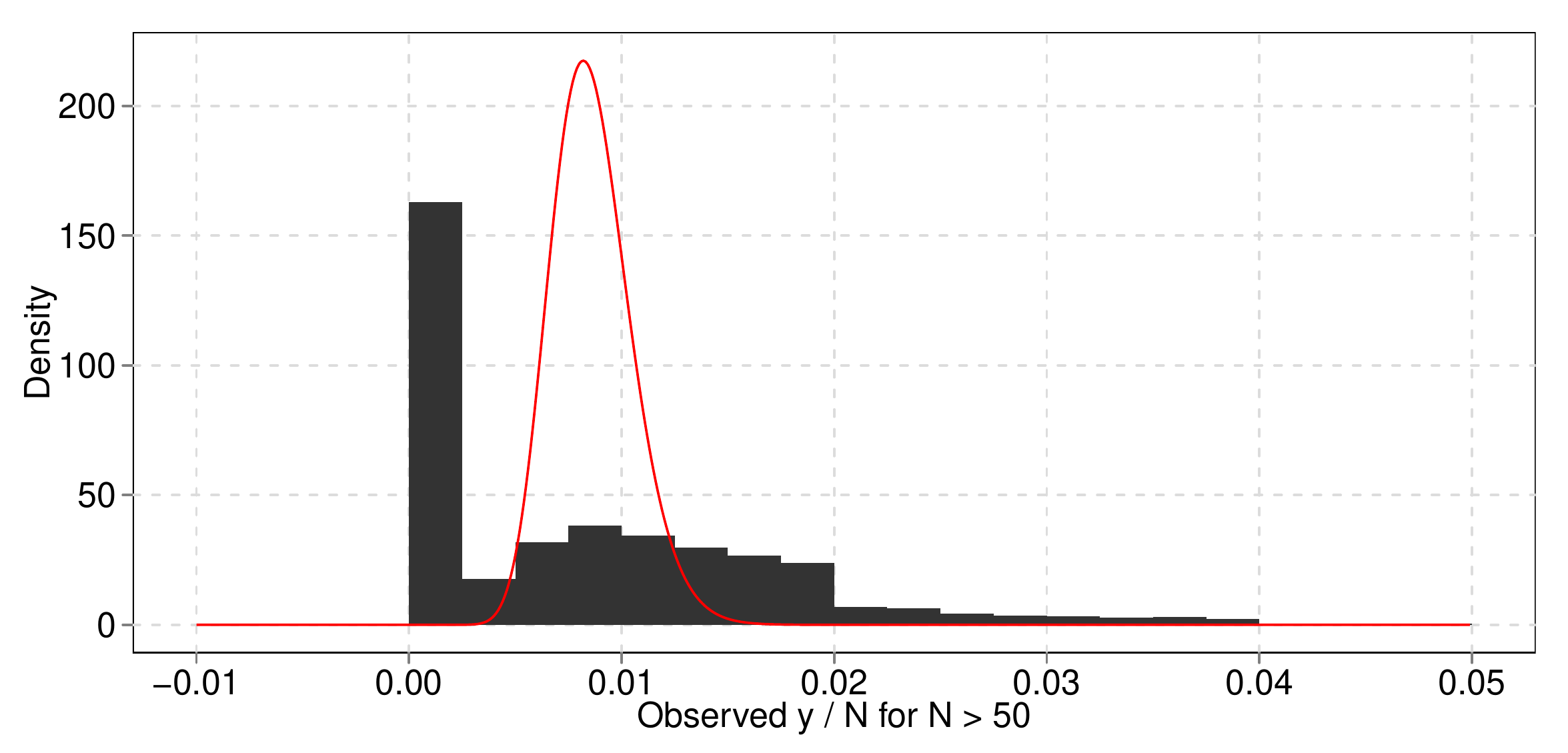}

\protect\caption{$\protect\priorestimate$ (red) and $y/N$ histogram for pairs with
$N\geq50$ in the middle bin. Note that the histogram and the red
distribution are not supposed to agree: the histogram is the red distribution,
plus Poisson noise.\label{fig:prior-example}}
\end{figure}

Figure \ref{fig:Unsmoothed-and-smooth} shows the fitted\textbf{ }$\E_{\hat{G}}\left(\conddist\right)$,
$\Var_{\hat{G}}\left(\conddist\right)$ as functions of $\baseestimate$.
The $\E_{\hat{G}}\left(\conddist\right)$, $\Var_{\hat{G}}\left(\conddist\right)$
curves are nicely behaved, but the variance is noisy. We smoothed
the curves and adjusted the $\priorestimate$ accordingly (we actually
smoothed $\E_{\hat{G}}\left(\log\conddist\right)$, $\Var_{\hat{G}}\left(\log\conddist\right)$
to make sure the $\priorestimate$ stayed positive).\textbf{ }Figure
\ref{fig:final-fitted-prior} shows the adjusted $\priorestimate$.
We used the adjusted $\priorestimate$ to get $\hat{\E}\left(\conddist\right)$,
$\hat{\Var}\left(\conddist\right)$, $\hat{\E}\left(\conddist,y\right)$
and $\hat{\Var}\left(\conddist,y\right)$, where
\[
\hat{\E}\left(\conddist\right)=\E_{\tilde{\prior}}\left(\conddist\right)
\]
and $\tilde{\prior}$ are the adjusted $\prior$ (the other expectations
and variances are defined similarly). Figure \ref{fig:prior-example}
shows $\priorestimate$ for the middle bin, along with $y/N$ for
the query-ad pairs with $N\geq50$. $\priorestimate$ is centered
at about the center of the $y/N$ histogram, but is narrower and smoother.
The $y/N$ histogram also has a big spike at $0$. The marginal $p$-histograms
fit well, indicating that while $\priorestimate$ does not have any
mass at $0$, it captures this spike when we add Poisson noise to
get the distribution of $y$. 

Evaluating these estimates is tricky since we do not know the true
CTRs. We cannot directly measure how well the point estimates match
$\param$, or check that $\hat{\Var}\left(\conddist,y\right)$ accurately
estimates the distance between $\param$ and $\E\left(\conddist,y\right).$
Instead, we evaluate our estimates by using them to make predictions
and predictive intervals. We randomly divided the impressions for
each query-ad pair into training ($90\%$) and test ($10\%$) sets,
and tried to predict the test data using the training data.

\begin{figure}
\includegraphics[width=1\textwidth]{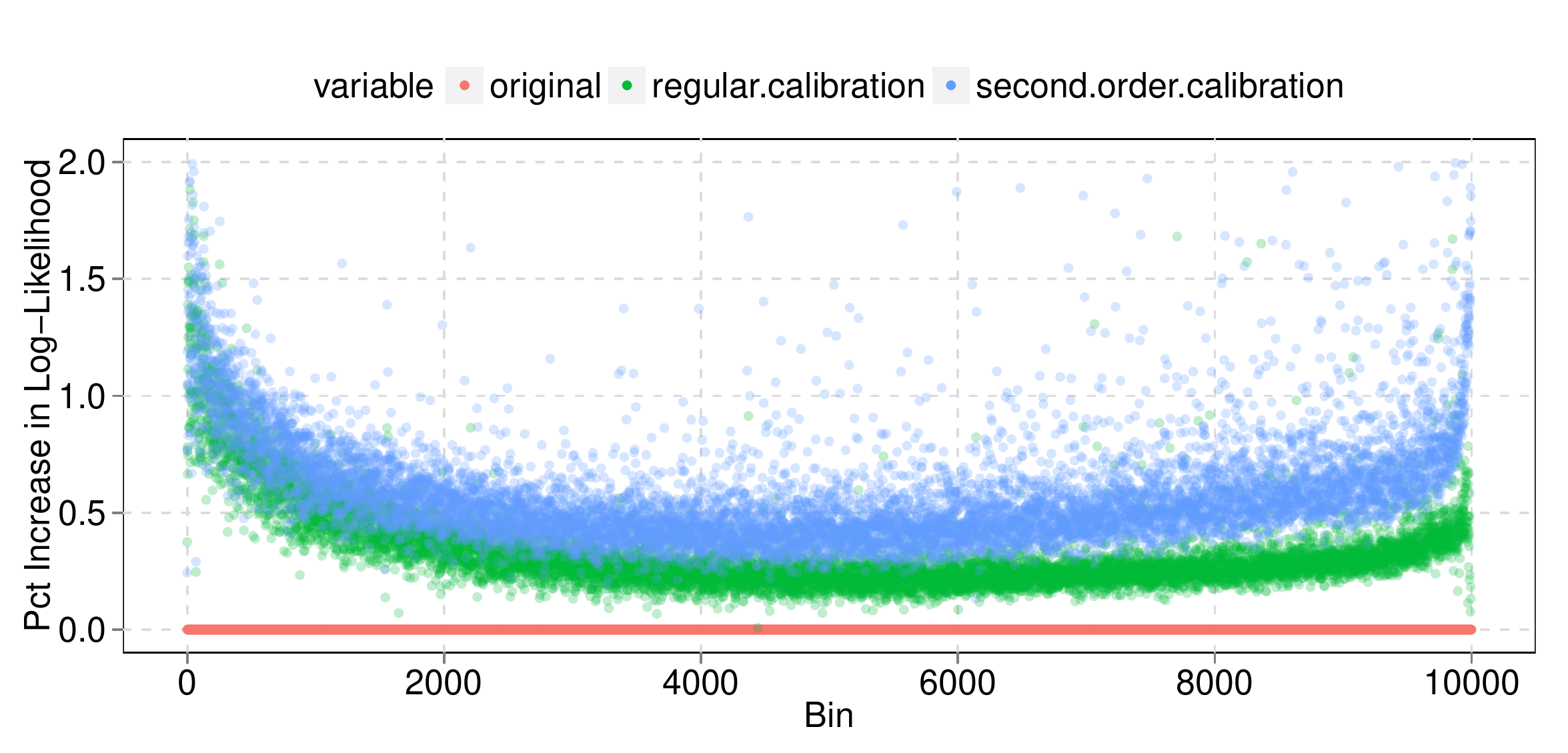}

\protect\caption{Percentage improvement in test set log-likelihood for $\protect\E\left(\protect\conddist\right)$
(green) and $\protect\E\left(\protect\conddist,y\right)$ (blue) over
$t$ (red baseline).\label{fig:LL-Percentage-improvement} }

\end{figure}

Second order calibration significantly improves our point estimates.
We judged the point estimates $\baseestimate$, $\E\left(\conddist\right)$
and $\E\left(\conddist,y\right)$ by their test set likelihood (under
the assumed Poisson noise). Figure\textbf{ }\ref{fig:LL-Percentage-improvement}
shows the improvement in test set likelihood for each bin. Both ordinary
and second order calibration increase test set likelihood in each
bin, and help more as we move away from the center. Overall, using
$\E\left(\conddist\right)$, the ordinary calibration estimate, instead
of $t$ increases the test set likelihood by $0.32\%$, and using
the second order calibration estimate $\E\left(\conddist,y\right)$
increases the test set likelihood by $0.63\%$. This is a substantial
increase, given the difficulty of CTR estimation and the degree to
which $t$ has been optimized. For comparison, using a constant, naive
estimate for all items (the average CTR in the middle bin) gives a
a test set likelihood \textbf{$13.6\%$ }lower than $t$.

\begin{figure}
\includegraphics[width=1\textwidth]{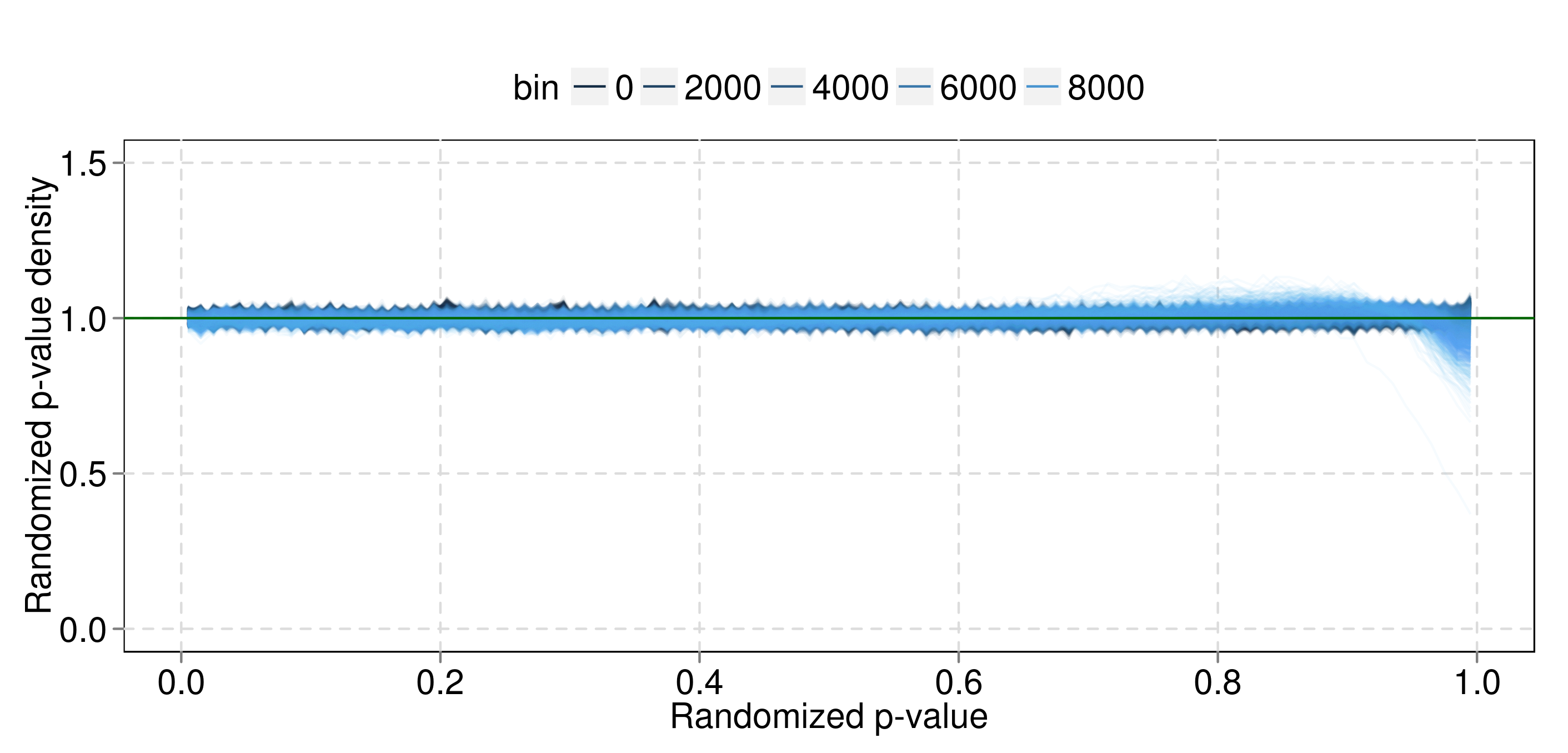}

\protect\caption{$p$-histogram densities for the predictive distribution within each
bin. The densities are mostly flat, indicating that our predictive
distributions mostly fit the test data.\label{fig:pred-rpval}}
\end{figure}

\begin{figure}
\includegraphics[width=1\textwidth]{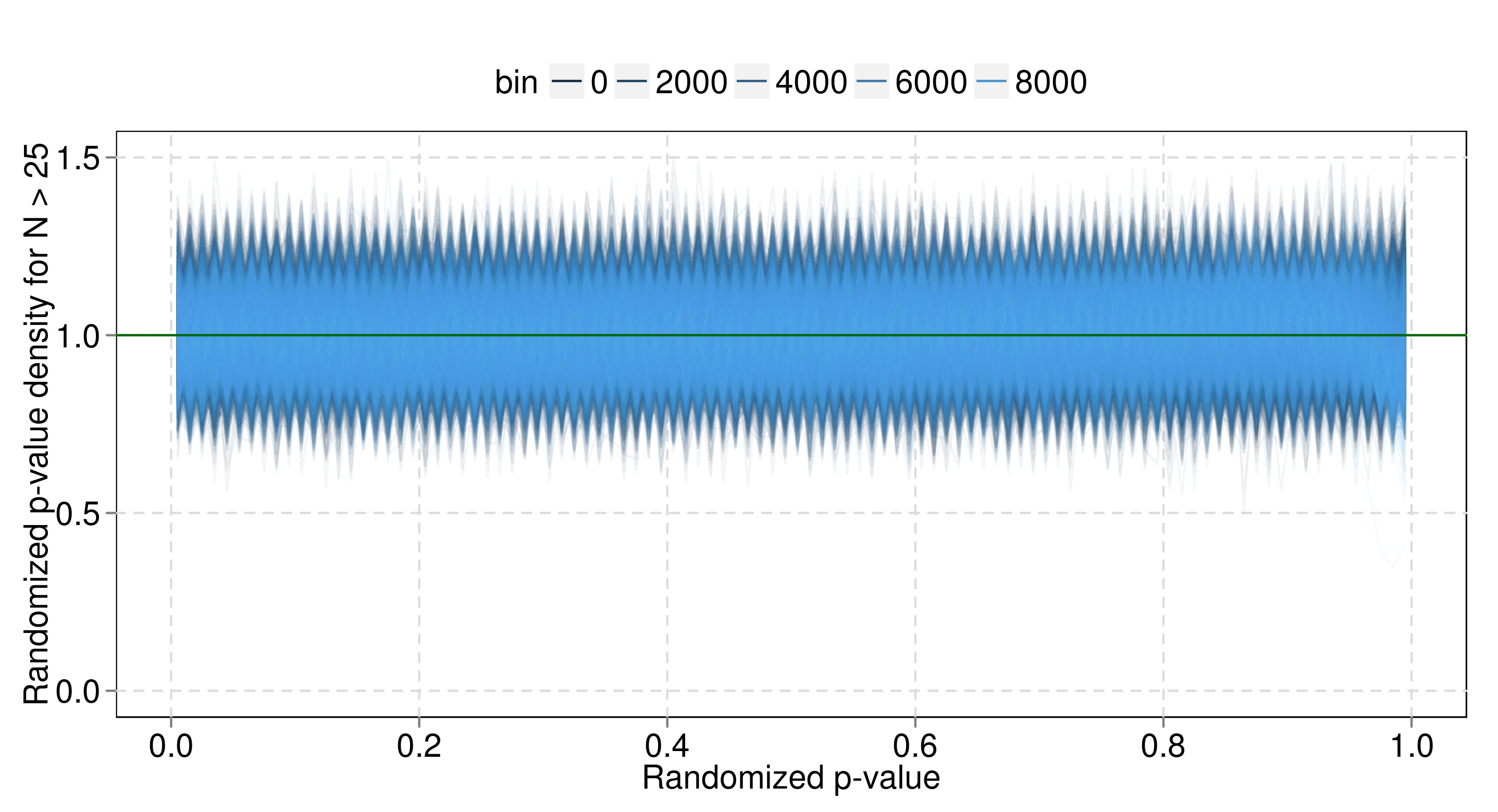}

\protect\caption{$p$-histogram densities for the predictive distribution within each
bin, restricted to query-ad pairs with $N\geq25$ in the test set.
The densities are mostly flat, indicating that our predictive distributions
mostly fit the test data.\label{fig:pred-rpval-big-n}}
\end{figure}

We used the fit of the predictive distributions to judge the accuracy
of $\hat{\Var}\left(\conddist,y\right)$. The predictive distributions
are wider when $\hat{\Var}\left(\conddist,y\right)$ is large and
shorter when $\hat{\Var}\left(\conddist,y\right)$ is small, so if
the predictive distributions match the data, $\hat{\Var}\left(\conddist,y\right)$
is probably measuring uncertainty well. We assessed the fit of the
predictive distributions with $p$-histograms like the one we used
to check the fit of the marginal distribution. Figures\textbf{ }\ref{fig:pred-rpval}
and \ref{fig:pred-rpval-big-n} show that the predictive distributions
fit well, both on the all the query-ad pairs, and those with \textbf{$N\geq25$}
in the test set.\textbf{ }The fit suggests that our $\Var\left(\conddist,y\right)$
estimates are accurate enough to be useful.

The fit of our point estimates and predictive distributions also suggests
that our Poisson model is approximately correct. To get the predictive
distributions right, we need to divide the variance of $y\given\baseestimate$
into signal ($\Var\left(\conddist\right)$) and noise ($y\given\param$).
If we underestimate the noise, for example, our estimate for $\Var\left(\conddist\right)$
will be too high. This means we will undershrink - our point estimates
and will be too close to $y/N$, and our predictive distributions
will be mis-centered. The fit of our predictive distributions indicates
that our model is putting about the right weight on signal and noise.

\begin{figure}
\includegraphics[width=1\textwidth]{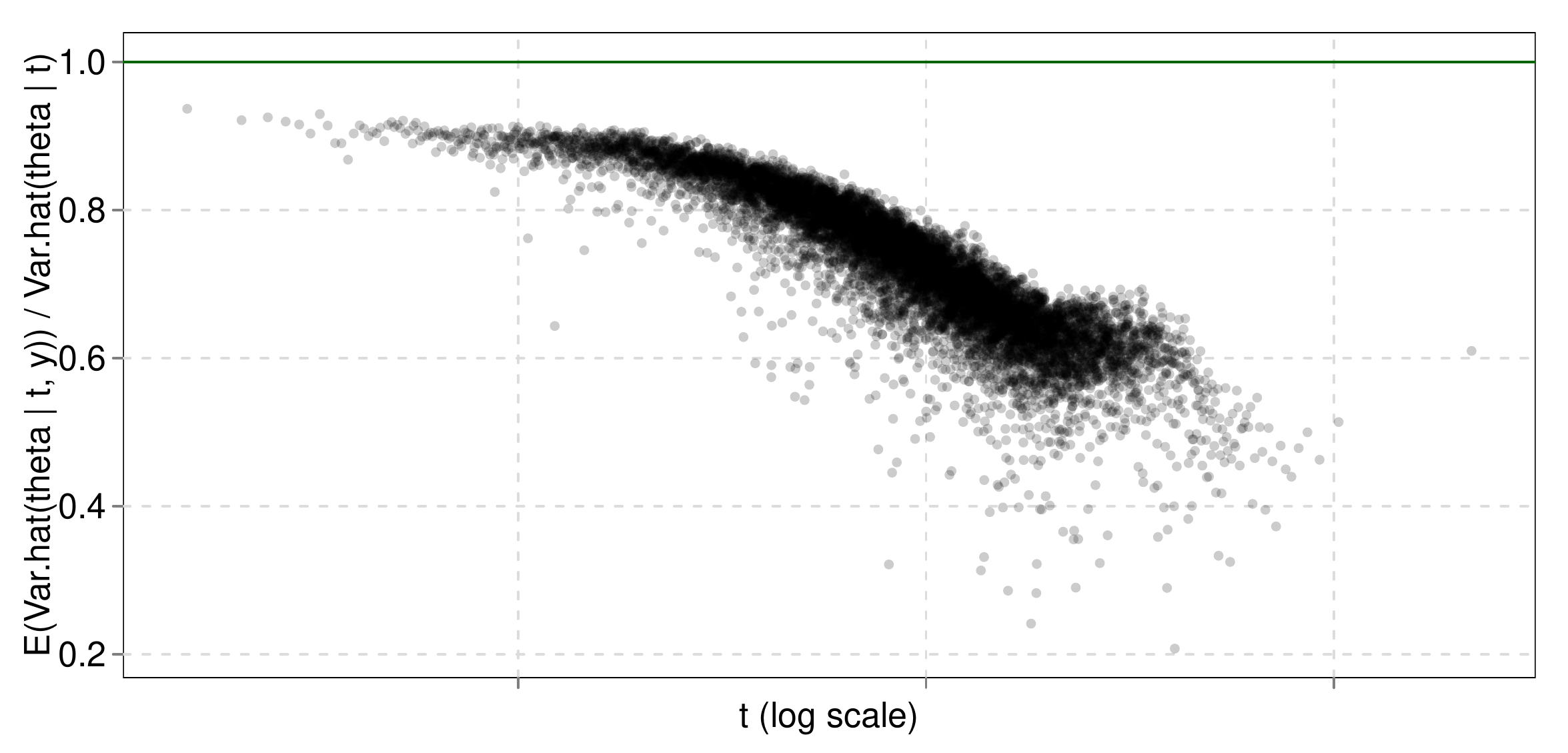}

\protect\caption{Plot of $\protect\E\left(\hat{\protect\Var}\left(\protect\conddist,y\right)\right)/\hat{\protect\Var}\left(\protect\conddist\right)$
for each $\protect\baseestimate$ bin. The drop in variance is biggest
in the high-$\protect\baseestimate$ bins, since query-ad pairs in
those bins have more clicks and thus more information.\label{fig:variance-gain}}
\end{figure}

Second order calibration gives us some interesting model-dependent
estimates of performance. For example, we can use the model to estimate
the fraction of the variance of $\param$ explained by $\baseestimate$:
\[
\hat{R}^{2}=1-\frac{\hat{\Var}\left(\conddist\right)}{\hat{\Var}\left(\param\right)}
\]
where 
\[
\hat{\Var}\left(\param\right)=\Var\left[\hat{\E}\left(\conddist\right)\right]+\E\left[\hat{\Var}\left(\conddist\right)\right]
\]
is the overall variance of $\param$, estimated using the conditional
variance formula and the fitted mean and variance of $\param$ in
each bin. For the CTR data, $\hat{R}^{2}$ was around $0.9$, indicating
that $\baseestimate$ is a strong predictor of $\param$. We could
use $\hat{R}^{2}$ to compare different candidates for $\baseestimate$
(\citet{Amini2009} use a similar metric based on a random effects
model). We can also use the model to estimate how much second order
calibration lowers the variance in our estimate of $\param$. Figure\textbf{
}\ref{fig:variance-gain} plots $\E\left(\hat{\Var}\left(\conddist,y\right)\right)/\hat{\Var}\left(\conddist\right)$
for each bin, where the expectation is weighted by $N$ in the test
set. Averaged over bins, this ratio was about $73\%$, which says
that second order calibration gives us a mean squared error about
$27\%$ less than ordinary calibration. The figure shows that the
variance reduction is biggest in the high-$\baseestimate$ bins; this
makes sense, since query-ad pairs in those bins have more clicks,
and thus more information about $\param$. Although these performance
estimates are model-dependent, and could be wrong because of model
misspecification, the fit and predictive distribution checks above
mean that the estimates are trustworthy enough to be interesting.

\subsection{Simulated CTR data}

Next, we tested our method on simulated CTR data that was based on
our real data set. We used the same $N$ and $\baseestimate$, generated
true $\param$ lognormally around $\baseestimate$ with some bias
and variance, and generated new clicks: 
\begin{eqnarray*}
\log\param_{i} & \sim & \mathcal{N}\left(\log\baseestimate_{i}+\delta,\sigma^{2}\right)\\
y_{i} & \sim & \Poisson\left(N_{i}\lambda_{i}\right).
\end{eqnarray*}
We used the same fitting method as for real data (so we had around
2-3 billion query-ad pairs), and looked at the results for different
values of $\delta$ and $\sigma$. To make computation easier, we
used every tenth bin instead of every bin.

\begin{figure}
\includegraphics[width=1\textwidth]{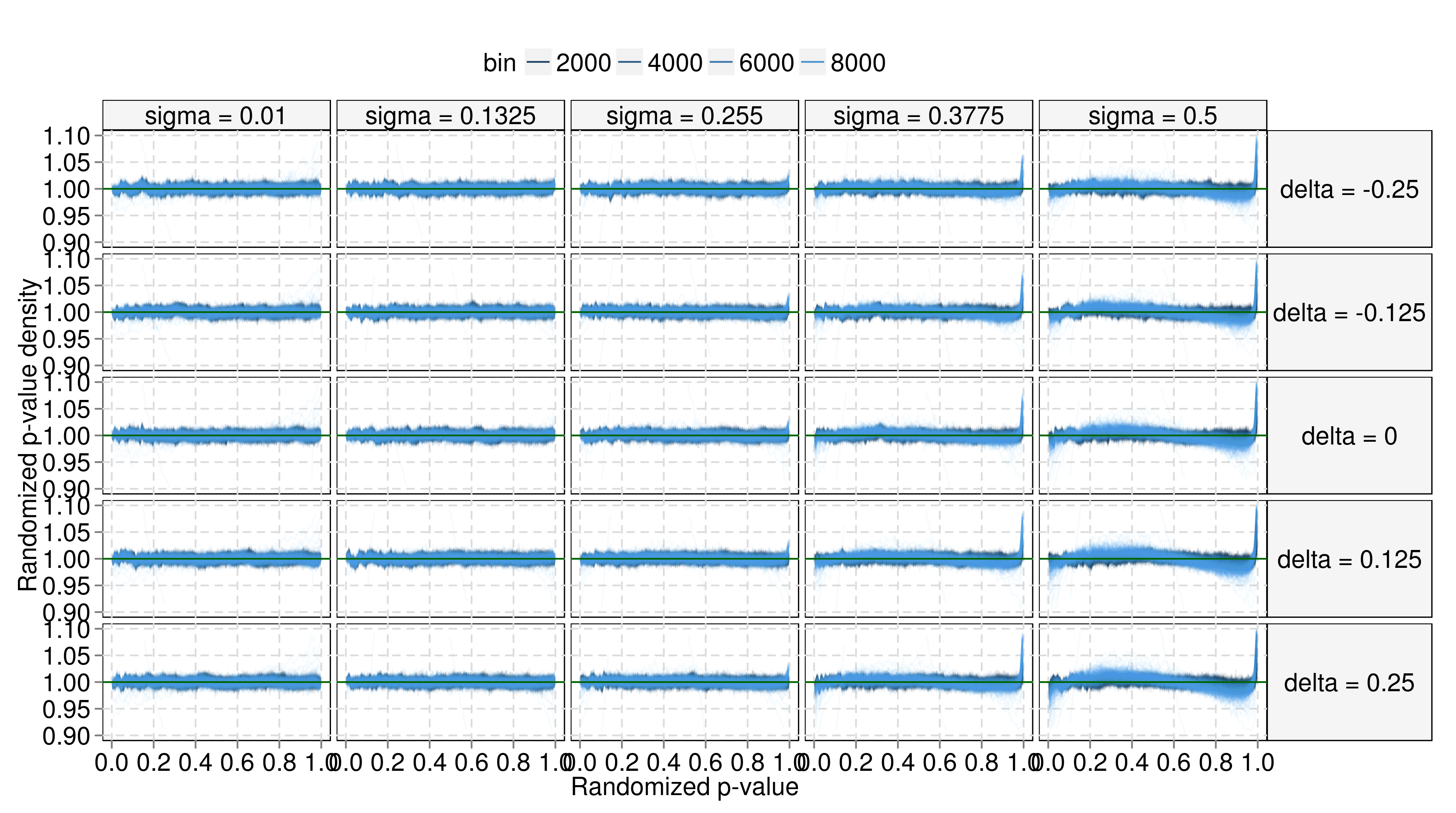}

\protect\caption{$p$-histogram densities for the marginal distribution within each
bin. The panels from show increasing $\sigma$ (top to bottom) and
$\delta$ (left to right).\label{fig:sim-marg-pval}}

\end{figure}

Figure\textbf{ }\ref{fig:sim-marg-pval} shows that a Gamma model
for $\prior$ fit the simulated data reasonably well - not surprising,
since the Gamma distribution approximates the lognormal distribution
well when $\sigma$ is small. The Gamma model doesn't quite fit the
data when $\sigma$ is large. We smoothed as before and calculated
$\hat{\E}\left(\conddist\right)$, $\hat{\Var}\left(\conddist\right)$,
$\hat{\E}\left(\conddist,y\right)$ and $\hat{\Var}\left(\conddist,y\right)$.

Our estimates are accurate for all the values of $\delta$ and $\sigma$
that we tried. We measured their accuracy directly, since we knew
the true $\conddist$ distributions. Figure\textbf{ }\ref{fig:Accuracy-plots-for}
shows that our estimates of $\E\left(\conddist\right)$, $\Var\left(\conddist\right)$,
$\E\left(\conddist,y\right)$ and $\Var\left(\conddist,y\right)$
were close to the corresponding true quantities. Calculating posterior
quantities in the true lognormal-Poisson model is computationally
expensive, so we found $\E\left(\conddist,y\right)$ and $\Var\left(\conddist,y\right)$
by approximating the lognormal with a Gamma distribution. This worked
better than approximation with a grid of point masses, and yielded
an accurate approximation for $\E\left(\conddist,y\right)$, but cannot
capture the heavy tails of the lognormal distribution and slightly
underestimated $\Var\left(\conddist,y\right)$. 
\begin{figure}
\subfloat[$\E\left(\conddist\right)$ and $\hat{\E}\left(\conddist\right)$
as functions of $t$.]{\includegraphics[width=0.85\columnwidth]{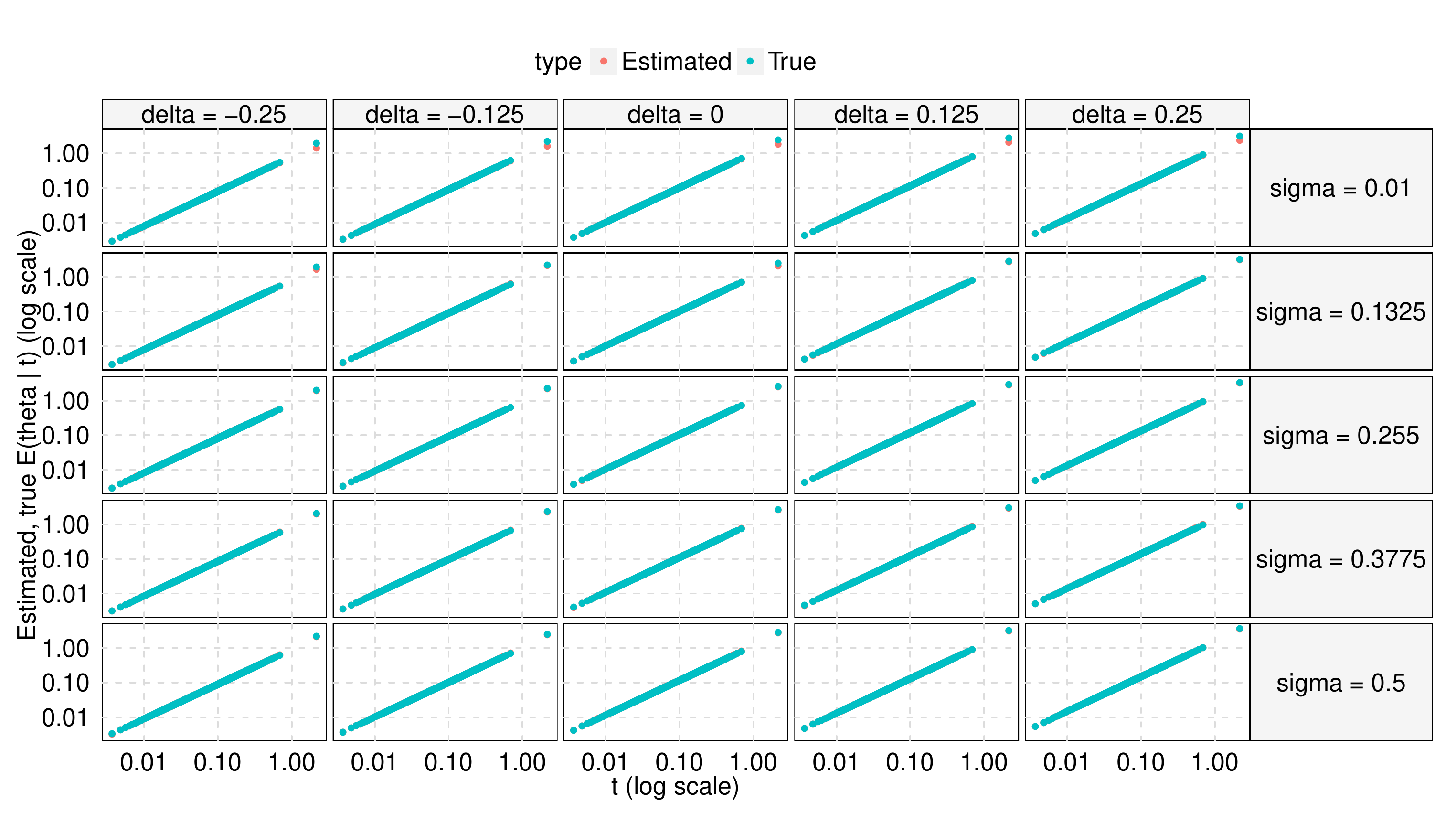}

}

\subfloat[$\Var\left(\conddist\right)$ and $\hat{\Var}\left(\conddist\right)$
as functions of $t$.]{\includegraphics[width=0.85\columnwidth]{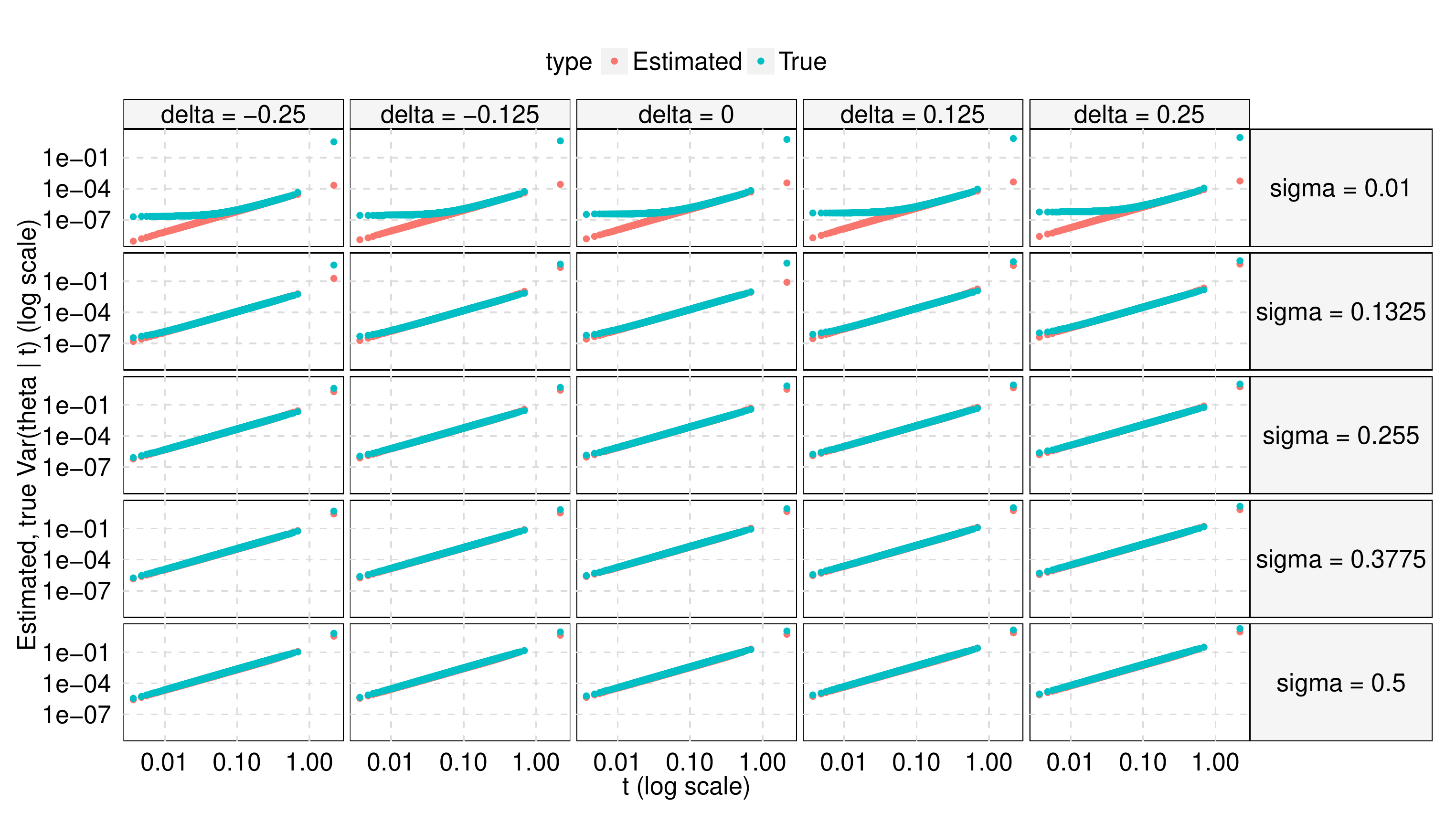}

}

\protect\caption{Accuracy plots for $\protect\E\left(\protect\conddist\right)$, $\protect\Var\left(\protect\conddist\right)$,
$\protect\E\left(\protect\conddist,y\right)$ and $\protect\Var\left(\protect\conddist,y\right)$.
For the first two, we plot the estimates and the true quantities.
For the second two, we plot the in-bin mean-squared-error between
the estimate and true quantities, scaled by $\protect\E\left(\protect\conddist\right)$
to put the different bins on the same scale. In each plot the panels
show increasing $\sigma$ (top to bottom) and $\delta$ (left to right).
Within each panel, each dot is a $t$-bin. All the estimated quantities
are pretty close to the true quantities.\label{fig:Accuracy-plots-for} }
\end{figure}
 
\begin{figure}
\ContinuedFloat

\subfloat[In-bin average $\left(\E\left(\conddist,y\right)-\hat{\E}\left(\conddist,y\right)\right)^{2}/\E\left(\theta\given t\right)^{2}$. ]{\includegraphics[width=0.85\columnwidth]{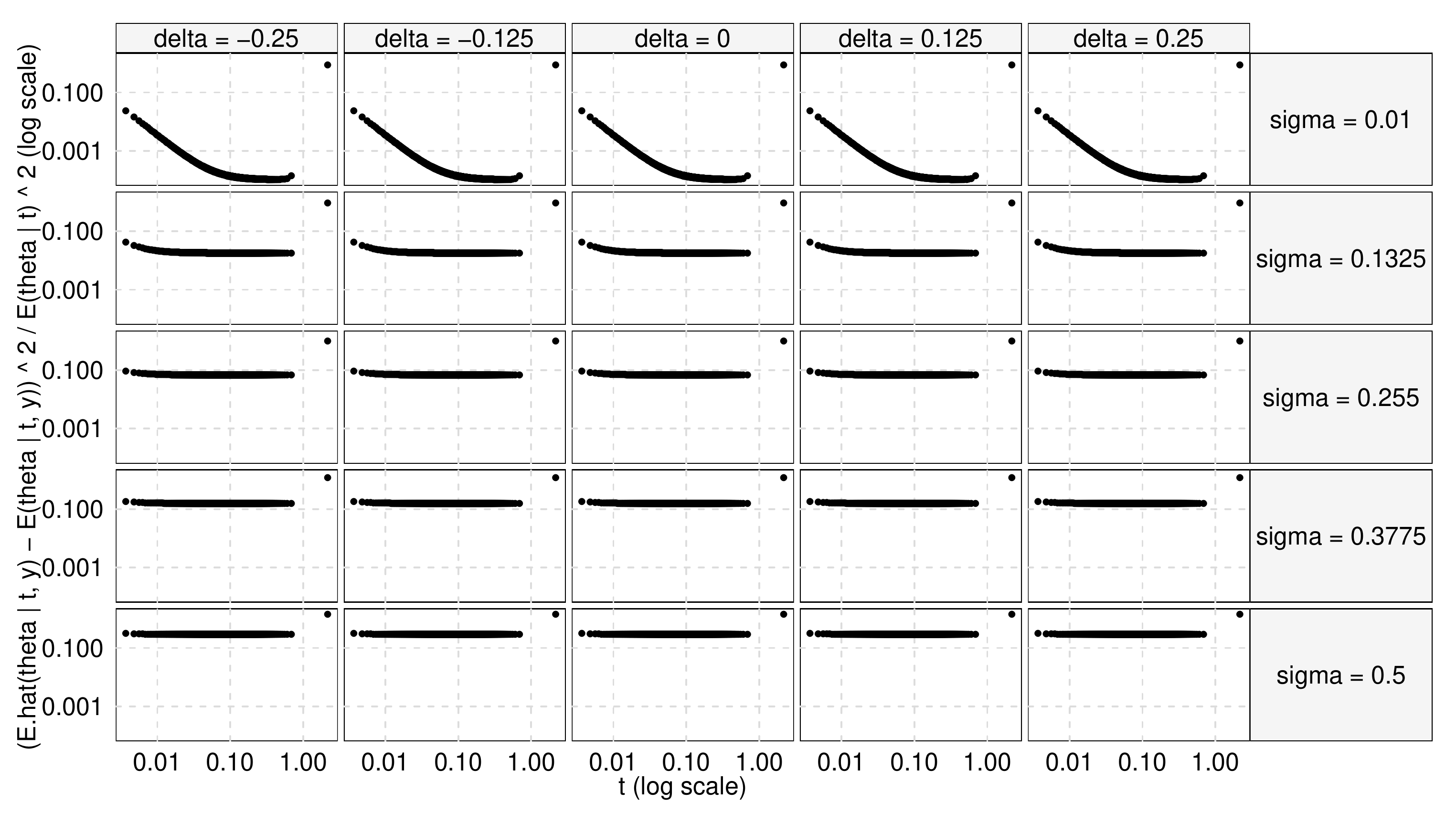}}

\subfloat[In-bin average $\left(\Var\left(\conddist,y\right)-\hat{\Var}\left(\conddist,y\right)\right)^{2}/\E\left(\theta\given t\right)^{4}$.]{\includegraphics[width=0.85\columnwidth]{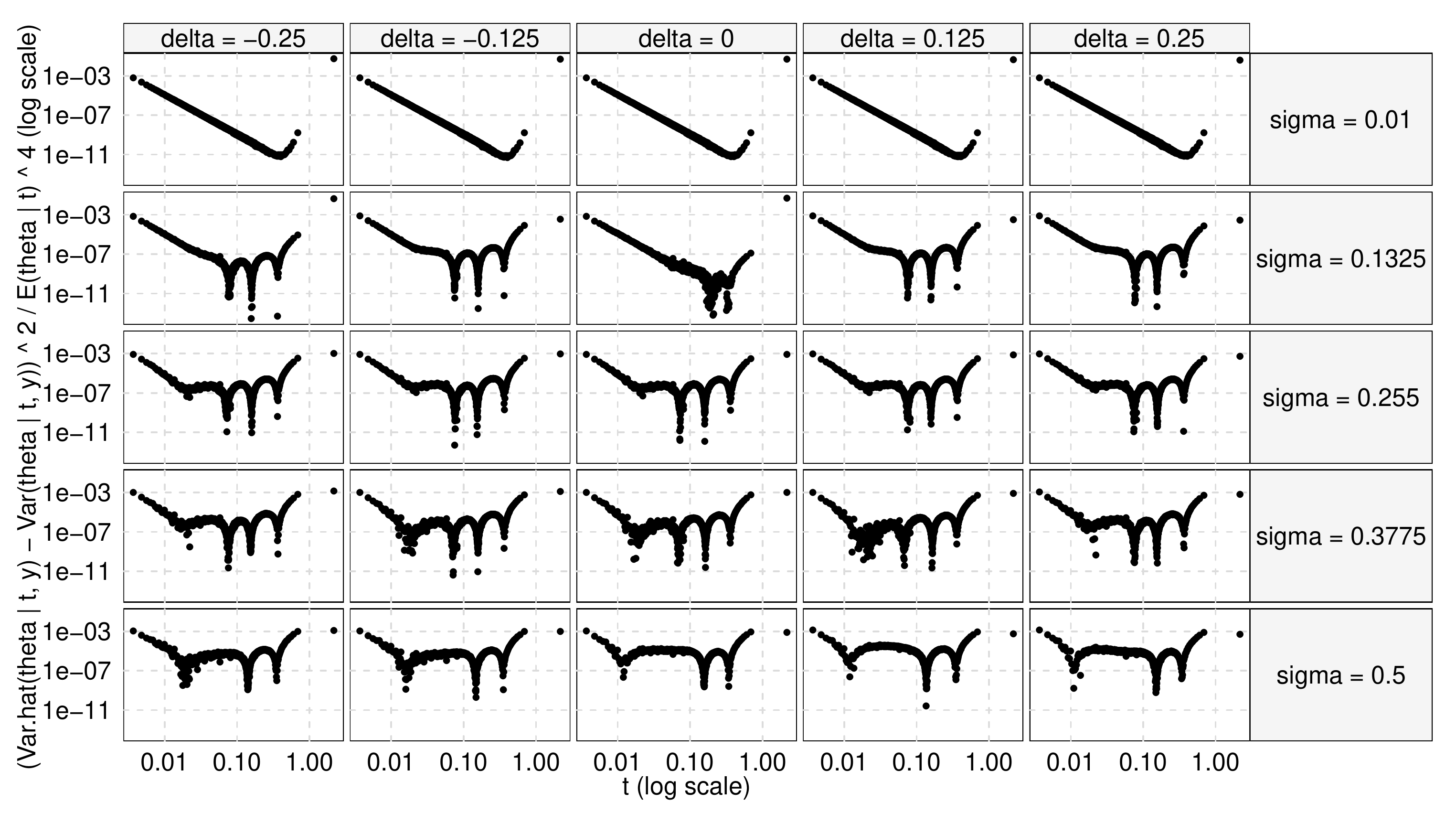}}

\protect\caption{(continued from previous page)}
\end{figure}

\clearpage

\begin{figure}
\includegraphics[width=1\columnwidth]{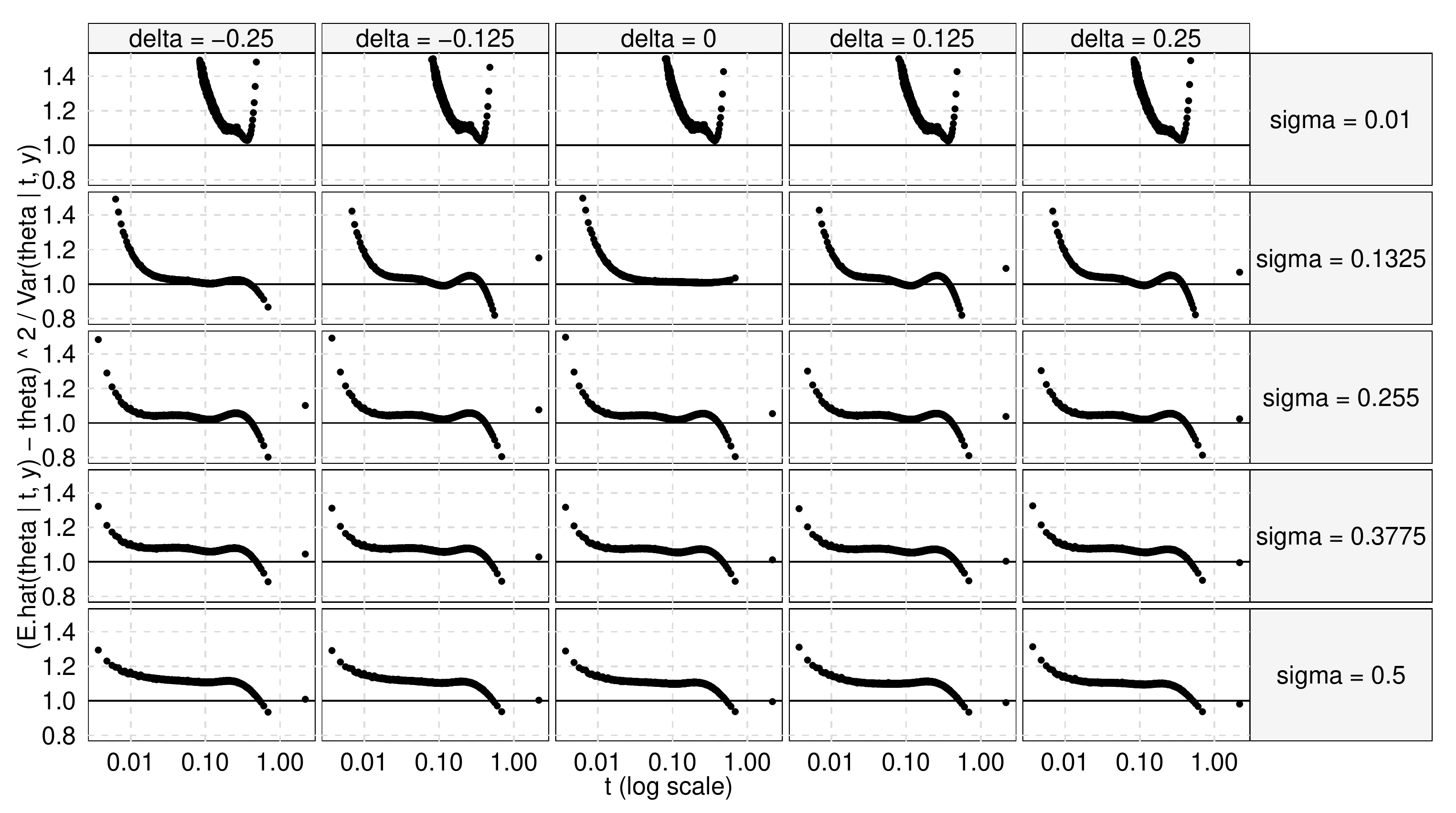}\protect\caption{In-bin averages of $\left(\hat{\protect\E}\left(\protect\conddist,y\right)-\theta\right)^{2}/\hat{\protect\Var}\left(\protect\conddist,y\right)$.
If our estimates were perfect, these would be $1$. Except when $\sigma$
is small, the ratios are close to $1$, but $5-10\%$ too high. This
indicates $\hat{\protect\Var}\left(\protect\conddist,y\right)$ is
underestimating the error of $\hat{\protect\E}\left(\protect\conddist,y\right)$;
our Gamma model cannot capture the heavy lognormal tails of the true
$\protect\conddist$ distribution.\label{fig:true-error-over-pred}}
\end{figure}

Figure \ref{fig:In-bin-mean-squared-error,} shows if this were a
real problem, second order calibration would not be perfect, but would
be good enough to be useful. It gives better point estimates - $\hat{\E}\left(\conddist,y\right)$
substantially improves on $\baseestimate$, $\hat{\E}\left(\conddist\right)$,
and $\E\left(\conddist\right)$, and estimates $\param$ almost as
well as our approximate $\E\left(\conddist,y\right)$. The improvement
is especially large when $\sigma$ is big, since the bigger $\sigma$
is, the more information is item-specific, and the more we can gain
by using it. Our variance estimates are also reasonably accurate.
On average, $\hat{\Var}\left(\conddist\right)$ is usually close to
$\left(\hat{E}\left(\conddist\right)-\theta\right)^{2}$, but is about
$5-10\%$ too small (Figure \ref{fig:true-error-over-pred}). This
is because the true distribution of $\conddist$ is lognormal, and
our Gamma model cannot capture its heavy tails. Our $p$-histograms
detect this misfit when $\sigma$ is large, but are not sensitive
enough to detect the misfit when $\sigma$ is small.
\begin{figure}
\includegraphics[width=1\columnwidth]{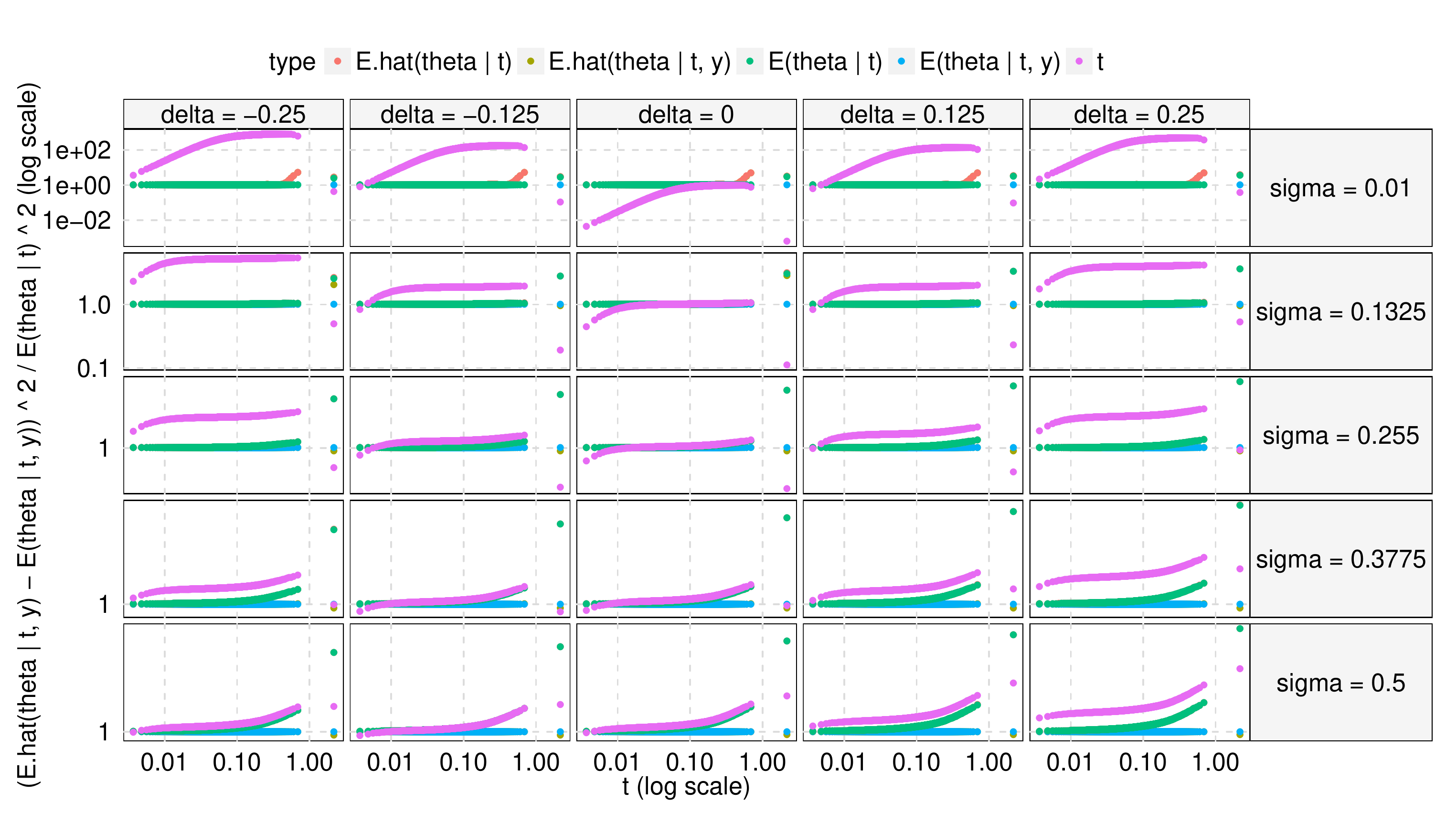}\protect\caption{In-bin mean-squared error, relative to the (approximated) $\protect\E\left(\protect\conddist,y\right)$.
The perfect estimator $\protect\E\left(\protect\conddist,y\right)$
has a relative error of $1$. $\hat{\protect\E}\left(\protect\conddist,y\right)$
almost always has relative error very close to $1$, so it is almost
as good as $\protect\E\left(\protect\conddist,y\right)$. $\protect\E\left(\protect\conddist\right)$
and $\hat{\protect\E}\left(\protect\conddist\right)$ are very close,
and perform worse than $\protect\E\left(\protect\conddist,y\right)$
when $\sigma$ is large. $t$ does badly, but actually outperforms
$\protect\E\left(\protect\conddist,y\right)$ when $\sigma$ and $\delta$
are both small - in that case, $\theta\approx t$, so $t$ is better
than $\protect\E\left(\protect\conddist,y\right)$, which only uses
$t$ through the bin.\label{fig:In-bin-mean-squared-error,}}

\end{figure}

\subsection{Simulated Normal data}

As a final illustration of our method, we consider a smaller simulated
normal data set with $10$ million items. Each item had a covariate
$\x\in\mathbb{R}^{10}$ with iid $\mathcal{N}\left(0,1\right)$ entries,
and a response $y\sim\mathcal{N}\left(\param,1\right)$. $\param$
was a quadratic function of $\x$, plus noise:

\[
\param=\x'\beta+\left(\x'\gamma\right)^{2}+\varepsilon,
\]
where $\varepsilon$ were drawn iid from a Laplace distribution with
variance $2$. The coefficient vectors $\beta$ and $\gamma$ each
had entries that were $0$ with probability $0.95$ and $\mathcal{N}\left(0,1\right)$
($\mathcal{N}\left(0,0.1^{2}\right)$ for $\gamma$) with probability
$0.05$. To get $\baseestimate$, we divided the data into five folds
and regressed $y$ onto $\boldsymbol{x}$ (with no interactions or
quadratic terms). Finally, we fit $\prior$ using a seven-component
normal mixture (using the R package ``mixfdr'' \citep{Muralidharan2010}),
and smoothed $\hat{\E}\left(\conddist\right)$ and $\hat{\Var}$$\left(\conddist\right)$. 

\begin{figure}
\includegraphics[width=1\textwidth]{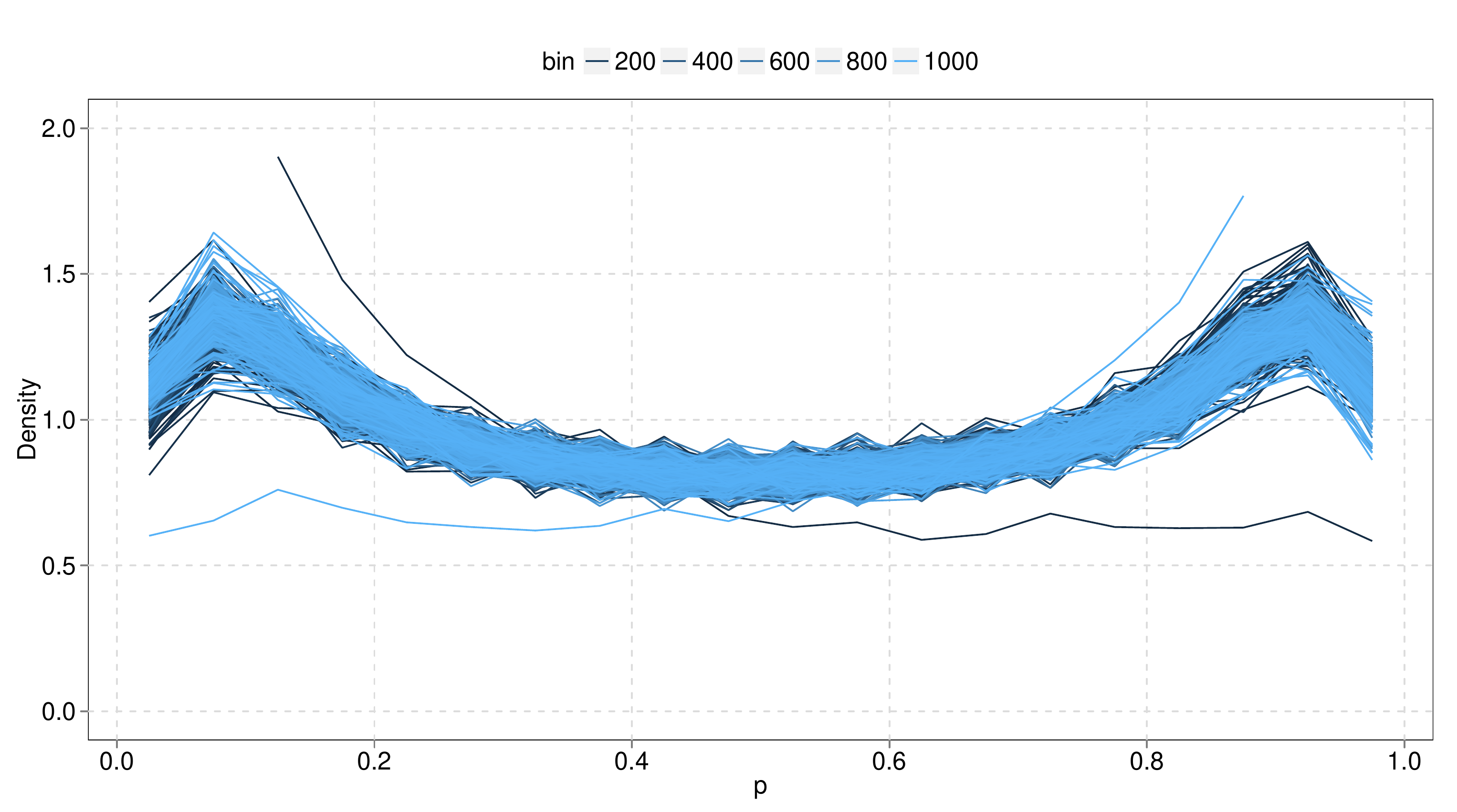}

\protect\caption{Marginal $p$-histograms for normal data. The ``smile'' indicates
that our model gives a light-tailed estimate for the distribution
of $y\protect\given\protect\baseestimate$. The left- and rightmost
bins are miscentered as well.\label{fig:normal-mpval}}

\end{figure}

Figure\textbf{ }\ref{fig:normal-mpval} shows the $p$-histograms
for our fitted model. Although the fit is decent in the center of
the distribution, we see many more low and high $p$s than our fitted
$y\given t$ densities predict. Our normal mixture model does not
capture the heavy tails of the distribution of $\conddist$. If this
were a real data set, the $p$-histograms would tell us our model
for $\prior$ doesn't fit, and we would refine it.

\begin{figure}
\includegraphics[width=1\textwidth]{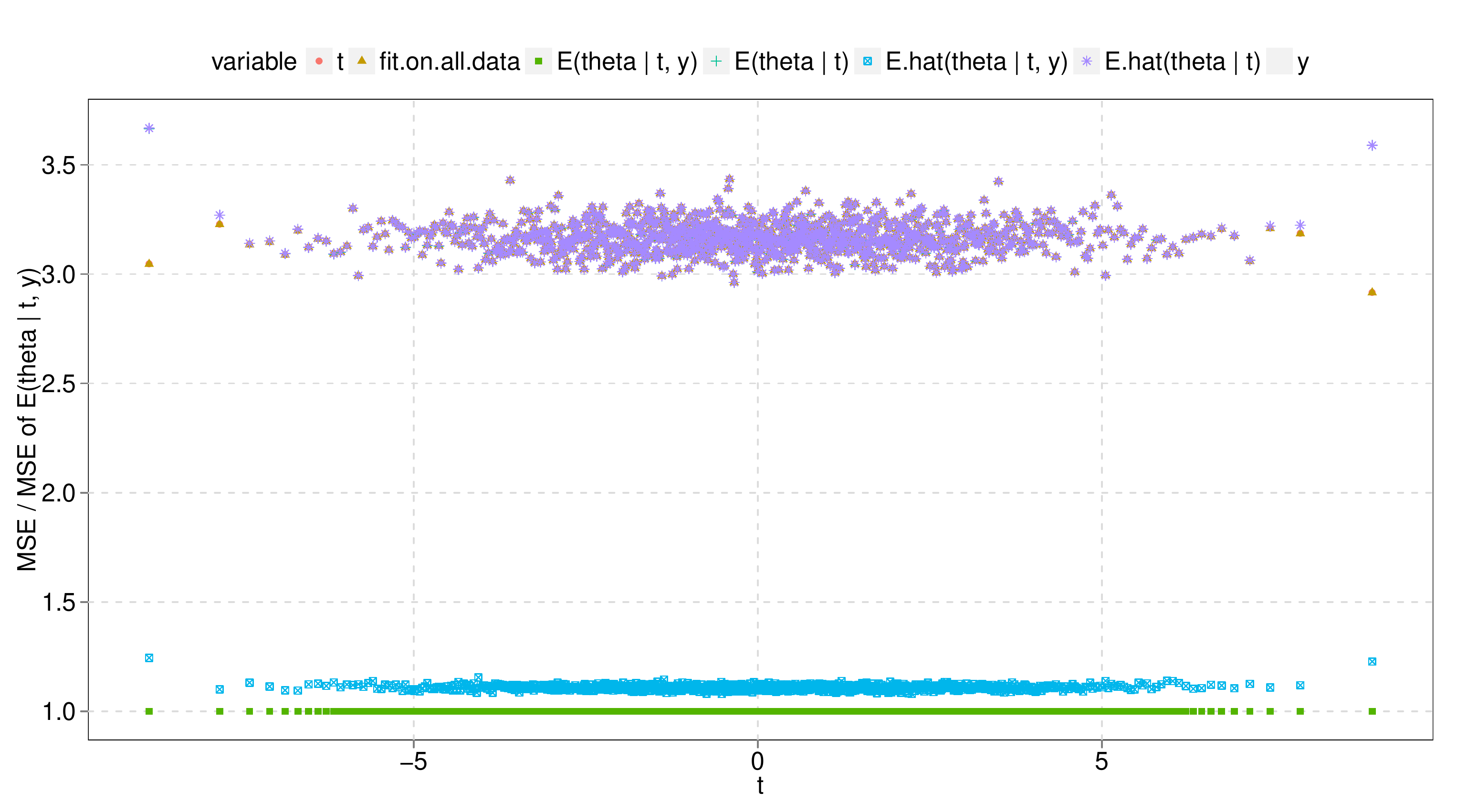}

\protect\caption{Mean squared error of the different estimators, relative to the mean
squared error of $\protect\E\left(\protect\conddist,y\right)$. $t$,
a regression on all the data (instead of dividing into folds), $\protect\E\left(\protect\conddist\right)$
and $\hat{\protect\E}\left(\protect\conddist\right)$ all perform
very similarly - their points are superimposed. $\hat{\protect\E}\left(\protect\conddist,y\right)$
(light blue) is about $11\%$ worse than $\protect\E\left(\protect\conddist,y\right)$.
\label{fig:normal-mpval-1}\label{fig:regret-normal}}
\end{figure}

\begin{figure}
\includegraphics[width=1\textwidth]{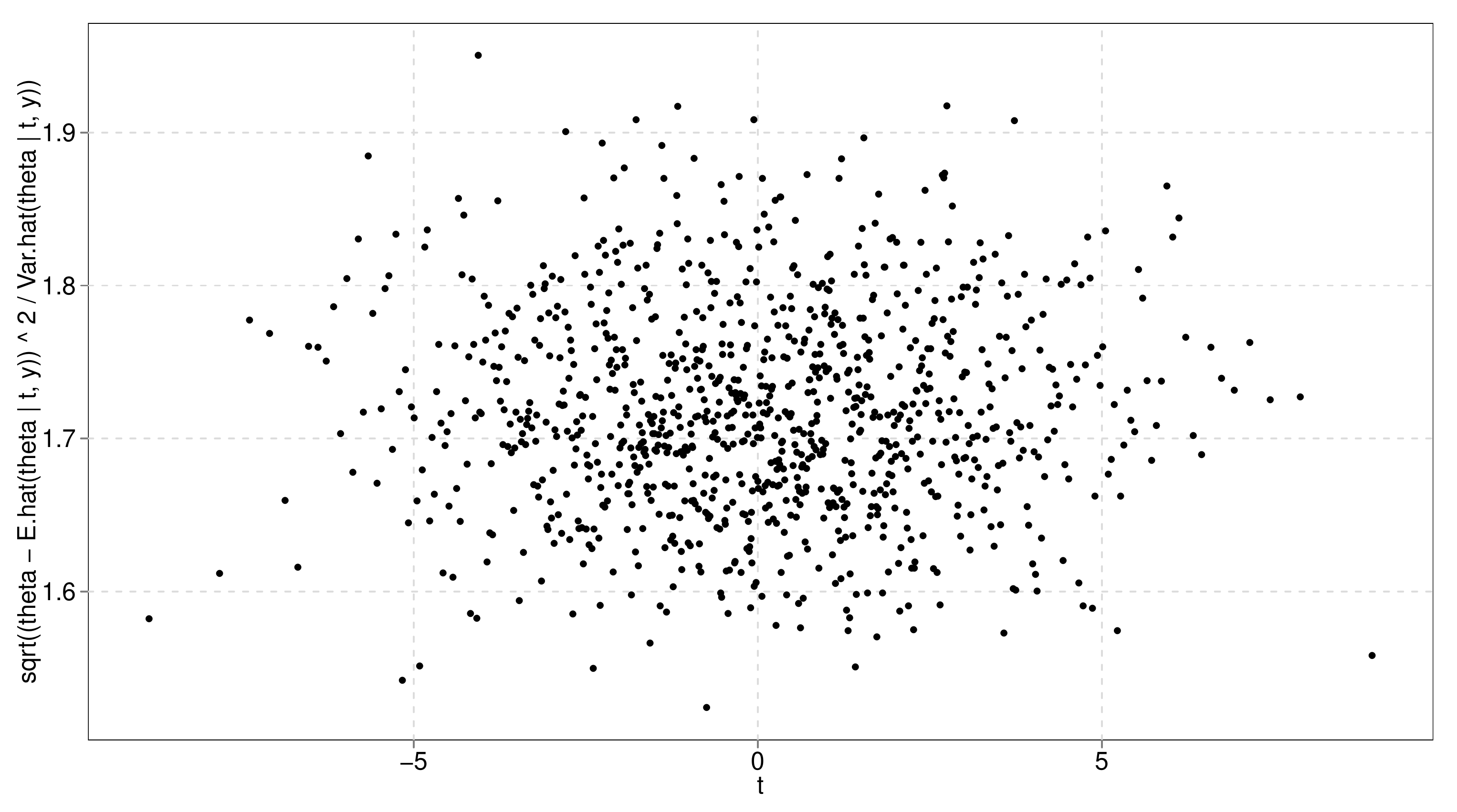}

\protect\caption{Square roots of the within-bin averages of $\left(\hat{\protect\E}\left(\protect\conddist,y\right)-\theta\right)^{2}/\hat{\protect\Var}\left(\protect\conddist,y\right)$.
On average, $\hat{\protect\E}\left(\protect\conddist,y\right)$ is
about $72\%$ further from $\theta$ than $\hat{\mbox{Sd}}\left(\protect\conddist,y\right)$.
\label{fig:error-ratio-normal}}
\end{figure}

It is interesting, though, to see how our flawed model performs. Figure\textbf{
}\ref{fig:regret-normal} shows that $\hat{\E}\left(\conddist,y\right)$
is a good, but not perfect point estimate. It estimates $\theta$
much more accurately than $t$ or $y$, and is about $11\%$ worse
than $\E\left(\conddist,y\right)$ (calculated using a fine grid approximation).
Figure \ref{fig:error-ratio-normal} shows that our light-tailed fit
makes our variance estimates too small - $\hat{\E}\left(\conddist,y\right)$
is, on average, about $72\%$ further from $\theta$ than $\hat{\Var}\left(\conddist,y\right)$
indicates. Theorem \ref{thm:error-bound} says that a better-fitting
model should perform better.

\section{Summary}

This paper considers second order calibration, a simple way to get
approximate posteriors from the output of an arbitrary black box estimation
method. The idea, which extends the usual idea of calibrating the
mean, is to approximate the distribution of $\param\given\x$ with
the distribution $\conddist$, and estimate the latter distribution
using the data. We give a five step procedure to estimate these quantities:
bin by $t$, estimate the distribution of $\conddist$ in each bin,
collect and if necessary smooth the estimates across bins, check the
fit, and use the estimates to calculate $\hat{\E}\left(\conddist\right)$,
$\hat{\Var}\left(\conddist\right)$, $\hat{\E}\left(\conddist,y\right)$
and $\hat{\Var}\left(\conddist,y\right)$. This is a reasonable thing
to do: if the distribution of $y\given\theta$ is Poisson or a continuous
natural exponential family, the data has enough information to estimate
$\E\left(\conddist\right)$, $\Var\left(\conddist\right)$, $\E\left(\conddist,y\right)$
and $\Var\left(\conddist,y\right)$ effectively. When applied to real
and simulated data, second order calibration improves point estimates
and gives useful accuracy estimates.

\subsection{Acknowledgements}

We thank many colleagues for helpful comments, suggestions and discussion:
Sugato Basu, Nick Chamandy, Brad Efron, Brendan McMahan, Donal McMahon,
Deirdre O'Brien, Daryl Pregibon and Tom Zhang.

\appendix

\section{Proofs\label{sec:Proofs}}

From now on, we work within a bin. We always condition on $\baseestimate$
(assumed constant in the bin), so we drop it to simplify notation.

\subsection{Posterior Cumulant Formulas}

In Section \ref{sub:Theoretical-Support}, we stated Robbins' formulas
for the posterior mean and variance of $\param$ when $f_{\theta}$
is $\Poisson\left(N\param\right)$. Similar formulas exist for higher
moments. Using Robbins' argument, it is easy to show that
\begin{equation}
\E\left(\param^{k}\given y\right)=\frac{\left[y\right]_{k}}{N^{k}}+\frac{1}{N^{k}}\frac{\Delta_{G}\left(y\right)}{f_{G}\left(y\right)}\label{eq:poission-moment}
\end{equation}
where $\left[y\right]_{k}=y\left(y-1\right)\ldots\left(y-k+1\right)$,
and $\Delta\left(y\right)=f_{G}\left(y+k\right)\left[y+k\right]_{k}-f_{G}\left(y\right)\left[y\right]_{k}$.
The two terms in the formula have a natural interpretation. The uniformly
minimum variance unbiased (UMVU) estimate of $\param^{k}$ is $\left[y\right]_{k}/N^{k}$,
so the first term in the formula is the UMVU estimate of $\theta^{k}$.
The second term is a correction that depends on the prior.

When $f_{\theta}$ is a continuous natural exponential family, it
is easier to work with cumulants than with moments. Let $\kappa_{k}$
be the $k$th cumulant of a distribution and let $p_{k}$ be the polynomial
that expresses the $k$th cumulant of a distribution in terms of the
first $k$ moments, so that for any distribution,
\[
\kappa_{k}=p_{k}\left(\mu_{1},\ldots,\mu_{k}\right)
\]
where $\mu_{j}$ is the $j$th moments. Let $\kappa_{k}\left(\param\given y\right)$
be the $k$th posterior cumulant of the $\param\given y$ distribution
(it depends on $G$, but we suppress this). Simple algebra shows that
if $f_{\theta}$ is a continuous natural exponential family with base
density $f_{0}$, then
\begin{equation}
\kappa_{k}\left(\param\given y\right)=p_{k}\left(\frac{f_{0}'\left(y\right)}{f_{0}\left(y\right)},\ldots,\left(-1\right)^{k}\frac{f_{0}^{\left(k\right)}\left(y\right)}{f_{0}\left(y\right)}\right)+p_{k}\left(\frac{f_{G}'\left(y\right)}{f_{G}\left(y\right)},\ldots,\frac{f_{G}^{\left(k\right)}\left(y\right)}{f_{G}\left(y\right)}\right).\label{eq:contfam-formula}
\end{equation}
The two terms in this formula have the same interpretation as the
two terms in equation \ref{eq:poission-moment}. The UMVU estimator
of $\param^{i}$ is $\left(-1\right)^{i}\frac{f_{0}^{\left(i\right)}}{f_{0}}$
\citep{Sharma1973}, so the first term plugs UMVU estimates of $\param,\ldots,\param^{k}$
into $p_{k}$ to estimate $\kappa_{k}$. The second term is a correction
that depends on the prior.

\subsection{Regularized Cumulant Estimators}

Equations \ref{eq:poission-moment} and \ref{eq:contfam-formula}
divide by $f_{G}\left(y\right)$. If our estimate $\hat{G}$ gives
a light-tailed $f_{\hat{G}}$, this division can make our posterior
moment and cumulant estimates behave badly. To avoid this, we follow
the approach of \citet{Zhang1997} and regularize our estimates: instead
of divding by $f_{\hat{G}}$, we divide by $\max\left(f_{\hat{G}},\rho\right)$,
where $\rho$ is a tuning parameter. This gives regularized estimators
\begin{eqnarray*}
\hat{\E}_{\rho}\left(\param^{k}\given y\right) & = & \frac{\left[y\right]_{k}}{N^{k}}+\frac{1}{N^{k}}\frac{\Delta_{\hat{G}}\left(y\right)}{\max\left(f_{\hat{G}}\left(y\right),\rho\right)}\\
\hat{\kappa}_{k,\rho}\left(\param\given y\right) & = & p_{k}\left(\frac{f_{0}'\left(y\right)}{f_{0}\left(y\right)},\ldots,\left(-1\right)^{k}\frac{f_{0}^{\left(k\right)}\left(y\right)}{f_{0}\left(y\right)}\right)+p_{k}\left(\frac{f_{\hat{G}}'\left(y\right)}{\max\left(f_{\hat{G}}\left(y\right),\rho\right)},\ldots,\frac{f_{\hat{G}}^{\left(k\right)}\left(y\right)}{\max\left(f_{\hat{G}}\left(y\right),\rho\right)}\right)
\end{eqnarray*}
instead of our original, unregularized estimators $\hat{\E}\left(\param^{k}\given y\right)$,
$\hat{\kappa}_{k}\left(\param\given y\right)$.

These regularized estimators guard against overshrinking. In the far
tail, the second term in each formula tends to zero, since $\max\left(f_{\hat{G}},\rho\right)$
becomes $\rho$ and the numerator$ $ of each ratio tends to zero.
That means that the correction term that depends on the prior disappears,
and our estimates reduce to frequentist estimators. This makes sense:
we don't know much about the prior in the far tail, so we shouldn't
deviate too much from the safe frequentist estimator. The regularized
estimators are similar in this respect to the limited translation
estimators introduced by \citet{Efron1971}.

\subsection{Proof of Theorem \ref{thm:error-bound}}

We prove Theorem \ref{thm:error-bound} by bounding the error in estimating
the posterior cumulants and moments in terms of the error in estimating
the marginal density. We first bound the error of the regularized
estimates, then use those bounds that to bound the error of the unregularized
estimates.
\begin{lem}
\label{lem:regularized-Poisson}The regularized Poisson moment estimator
has error at most
\[
\left\Vert \hat{\E}_{\rho}\left(\theta^{k}\given y\right)-\E\left(\theta^{k}\given y\right)\right\Vert \leq\frac{C_{G,\rho}}{N^{k}}\left(\left\Vert \left(f_{\hat{G}}-f_{G}\right)^{2}\right\Vert ^{\frac{1}{2}}+\left\Vert \left(f_{\hat{G}}\left(y+k\right)-f_{G}\left(y+k\right)\right)^{2}\right\Vert ^{\frac{1}{2}}\right)+\frac{D_{G,\rho}}{N^{k}}
\]
where $C_{G,\rho}$, $D_{G,\rho}$ only depend on $G$ and $\rho$:
$C_{G,\rho}=\frac{1}{\rho^{2}}\left\Vert \Delta_{G}^{2}\right\Vert ^{\frac{1}{2}}+\frac{1}{\rho}\left\Vert \left[y\right]_{k}^{2}\right\Vert ^{\frac{1}{2}}+\frac{1}{\rho}\left\Vert \left[y+k\right]_{k}^{2}\right\Vert ^{\frac{1}{2}}$
and $D_{G,\rho}=\left\Vert \left(\frac{\Delta_{G}\left(y\right)}{f_{G}\left(y\right)}\right)\left(1-\frac{f_{G}\left(\rho\right)}{\rho}\right)_{+}\right\Vert $.\end{lem}
\begin{proof}
We first bound $\left\Vert \hat{\E}_{\rho}\left(\theta^{k}\given y\right)-\E\left(\theta^{k}\given y\right)\right\Vert $:
\[
\left\Vert \hat{\E}_{\rho}\left(\theta^{k}\given y\right)-\E\left(\theta^{k}\given y\right)\right\Vert \leq N^{-k}\left\Vert \left(\frac{\Delta_{\hat{G}}\left(y\right)}{f_{\hat{G}}\left(y\right)\vee\rho}-\frac{\Delta_{G}\left(y\right)}{f_{G}\left(y\right)\vee\rho}\right)\right\Vert +N^{-k}\left\Vert \left(\frac{\Delta_{G}\left(y\right)}{f_{G}\left(y\right)\vee\rho}-\frac{\Delta_{G}\left(y\right)}{f_{G}\left(y\right)}\right)\right\Vert 
\]
The second term is $D_{G,\rho}$. We bound the first term using Cauchy-Schwartz
and the triangle inequality:
\begin{eqnarray*}
\left\Vert \left(\frac{\Delta_{\hat{G}}\left(y\right)}{f_{\hat{G}}\left(y\right)\vee\rho}-\frac{\Delta_{G}\left(y\right)}{f_{G}\left(y\right)\vee\rho}\right)\right\Vert  & \leq & \left\Vert \frac{\Delta_{G}}{f_{G}\vee\rho}\left(\frac{f_{G}\vee\rho-f_{\hat{G}}\vee\rho}{f_{\hat{G}}\vee\rho}\right)\right\Vert +\left\Vert \left(\frac{\Delta_{G}-\Delta_{\hat{G}}}{f_{\hat{G}}\vee\rho}\right)\right\Vert \\
 & \leq & \frac{1}{\rho^{2}}\left\Vert \Delta_{G}^{2}\right\Vert ^{\frac{1}{2}}\left\Vert \left(f_{G}-f_{\hat{G}}\right)^{2}\right\Vert ^{\frac{1}{2}}\\
 &  & +\frac{1}{\rho}\left\Vert \left[y\right]_{k}^{2}\right\Vert ^{\frac{1}{2}}\left\Vert \left(f_{G}-f_{\hat{G}}\right)^{2}\right\Vert ^{\frac{1}{2}}\\
 &  & +\frac{1}{\rho}\left\Vert \left[y+k\right]_{k}^{2}\right\Vert ^{\frac{1}{2}}\left\Vert \left(f_{G}\left(y+k\right)-f_{\hat{G}}\left(y+k\right)\right)^{2}\right\Vert ^{\frac{1}{2}}.
\end{eqnarray*}
\end{proof}
\begin{lem}
\label{lem:regularized-continuous}The regularized posterior cumulant
estimator for continuous natural exponential families has error at
most
\begin{eqnarray*}
\left\Vert \hat{\kappa}_{k,\rho}\left(\param\given y\right)-\kappa_{k}\left(\param\given y\right)\right\Vert  & \leq & C_{G,\hat{G},\rho}\left\Vert \left(\sum_{i=0}^{k}\left(f_{G}^{\left(i\right)}-f_{\hat{G}}^{\left(i\right)}\right)^{2}\right)^{\frac{1}{2}}\right\Vert \\
 &  & +\frac{1}{\rho}\left\Vert p_{k}\left(\frac{f_{G}'}{f_{G}\vee\rho},\ldots,\frac{f_{G}^{\left(k\right)}}{f_{G}\vee\rho}\right)\right\Vert ^{\frac{1}{2}}\left\Vert \left(f_{G}-f_{\hat{G}}\right)^{2}\right\Vert ^{\frac{1}{2}}\\
 &  & +\left\Vert p_{k}\left(\frac{f_{G}'}{f_{G}\vee\rho},\ldots,\frac{f_{G}^{\left(k\right)}}{f_{G}\vee\rho}\right)-p_{k}\left(\frac{f_{G}'}{f_{G}},\ldots,\frac{f_{G}^{\left(k\right)}}{f_{G}}\right)\right\Vert 
\end{eqnarray*}
\end{lem}
\begin{proof}
We have
\begin{eqnarray*}
\left\Vert \hat{\kappa}_{k,\rho}\left(\param\given y\right)-\kappa_{k}\left(\param\given y\right)\right\Vert  & \le & \left\Vert p_{k}\left(\frac{f_{\hat{G}}'}{f_{\hat{G}}\vee\rho},\ldots,\frac{f_{\hat{G}}^{\left(k\right)}}{f_{\hat{G}}\vee\rho}\right)-p_{k}\left(\frac{f_{G}'}{f_{G}\vee\rho},\ldots,\frac{f_{G}^{\left(k\right)}}{f_{G}\vee\rho}\right)\right\Vert \\
 &  & +\left\Vert p_{k}\left(\frac{f_{G}'}{f_{G}\vee\rho},\ldots,\frac{f_{G}^{\left(k\right)}}{f_{G}\vee\rho}\right)-p_{k}\left(\frac{f_{G}'}{f_{G}},\ldots,\frac{f_{G}^{\left(k\right)}}{f_{G}}\right)\right\Vert .
\end{eqnarray*}
To bound the first term, we write $p_{k}\left(\frac{f'}{f},\ldots,\frac{f^{\left(k\right)}}{f}\right)=\frac{1}{f^{k}}H\left(f',\ldots,f^{\left(k\right)}\right)$
where $H$ is a polynomial of degree $k$; we can do this since every
term in $p_{k}$ has degree $k$. Let $B$ be the box $\prod_{i=1}^{k}\left[-\max\left(\left\Vert f_{G}^{\left(i\right)}\right\Vert _{\infty},\left\Vert f_{\hat{G}}^{\left(i\right)}\right\Vert _{\infty}\right),\max\left(\left\Vert f_{G}^{\left(i\right)}\right\Vert _{\infty},\left\Vert f_{\hat{G}}^{\left(o\right)}\right\Vert _{\infty}\right)\right]$,
and let $C=\sup_{B}\left\Vert \nabla H\right\Vert _{2}$ be the maximum
of the $\ell^{2}$ norm of the gradient over $B$. $C$ only depends
on $G$, $\hat{G}$ through $\left\Vert f_{G}^{\left(i\right)}\right\Vert _{\infty}$
and $\left\Vert f_{\hat{G}}^{\left(i\right)}\right\Vert _{\infty}$.
Then the first term is bounded by
\begin{eqnarray*}
 &  & \left\Vert \frac{1}{\left(f_{\hat{G}}\vee\rho\right)^{k}}H\left(f_{\hat{G}}',\ldots,f_{\hat{G}}^{\left(k\right)}\right)-\frac{1}{\left(f_{G}\vee\rho\right)^{k}}H\left(f_{G}',\ldots,f_{G}^{\left(k\right)}\right)\right\Vert \\
 & = & \left\Vert \frac{1}{\left(f_{\hat{G}}\vee\rho\right)^{k}}H\left(f_{\hat{G}}',\ldots,f_{\hat{G}}^{\left(k\right)}\right)-\frac{1}{\left(f_{G}\vee\rho\right)^{k}}H\left(f_{G}',\ldots,f_{G}^{\left(k\right)}\right)\right\Vert \\
 & \leq & \frac{1}{\rho^{k}}\left\Vert H\left(f_{\hat{G}}',\ldots,f_{\hat{G}}^{\left(k\right)}\right)-H\left(f_{G}',\ldots,f_{G}^{\left(k\right)}\right)\right\Vert +\frac{1}{\rho}\left\Vert \frac{H\left(f_{G}',\ldots,f_{G}^{\left(k\right)}\right)}{\left(f_{G}\vee\rho\right)^{k}}\left(f_{G}\vee\rho-f_{\hat{G}}\vee\rho\right)\right\Vert \\
 & \leq & \frac{C}{\rho^{k}}\left\Vert \left(\sum\left(f_{G}^{\left(i\right)}-f_{\hat{G}}^{\left(i\right)}\right)^{2}\right)^{\frac{1}{2}}\right\Vert +\frac{1}{\rho}\left\Vert \left(\frac{H\left(f_{G}',\ldots,f_{G}^{\left(k\right)}\right)}{f_{G}^{k}}\right)^{2}\right\Vert ^{\frac{1}{2}}\left\Vert \left(f_{G}-f_{\hat{G}}\right)^{2}\right\Vert ^{\frac{1}{2}}
\end{eqnarray*}
\end{proof}
\begin{lem}
\label{lem:unregularized-Poisson}The unregularized Poisson moment
estimator has error at most

\[
\left\Vert \hat{\E}\left(\theta^{k}\given y\right)-\E\left(\theta^{k}\given y\right)\right\Vert \leq N^{-k}\inf_{\rho}\left[C_{G,\rho}\left(\left\Vert \left(f_{\hat{G}}-f_{G}\right)^{2}\right\Vert ^{\frac{1}{2}}+\left\Vert \left(f_{\hat{G}}\left(y+k\right)-f_{G}\left(y+k\right)\right)^{2}\right\Vert ^{\frac{1}{2}}\right)+D_{G,\rho}+D_{\hat{G},\rho}\right]
\]
\end{lem}
\begin{proof}
For any $\rho$,
\[
\left\Vert \hat{\E}\left(\theta^{k}\given y\right)-\E\left(\theta^{k}\given y\right)\right\Vert \leq\left\Vert \hat{\E}_{\rho}\left(\theta^{k}\given y\right)-\hat{\E}\left(\theta^{k}\given y\right)\right\Vert +\left\Vert \hat{\E}_{\rho}\left(\theta^{k}\given y\right)-\E\left(\theta^{k}\given y\right)\right\Vert .
\]
The first term is $N^{-k}D_{\hat{G},\rho}$, and we can bound the
second term using Lemma \ref{lem:regularized-Poisson}. Taking the
minimum over $\rho$ finishes the proof.\end{proof}
\begin{lem}
\label{lem:unregularized-continuous}The unregularized posterior cumulant
estimator for continuous natural exponential families has error at
most

\begin{eqnarray*}
\left\Vert \hat{\kappa}_{k,\rho}\left(\param\given y\right)-\kappa_{k}\left(\param\given y\right)\right\Vert  & \leq & \inf_{\rho}\Biggl[C_{G,\hat{G},\rho}\left\Vert \left(\sum_{i=0}^{k}\left(f_{G}^{\left(i\right)}-f_{\hat{G}}^{\left(i\right)}\right)^{2}\right)^{\frac{1}{2}}\right\Vert \\
 &  & +\frac{1}{\rho}\left\Vert p_{k}\left(\frac{f_{G}'}{f_{G}\vee\rho},\ldots,\frac{f_{G}^{\left(k\right)}}{f_{G}\vee\rho}\right)\right\Vert ^{\frac{1}{2}}\left\Vert \left(f_{G}-f_{\hat{G}}\right)^{2}\right\Vert ^{\frac{1}{2}}\\
 &  & +\left\Vert p_{k}\left(\frac{f_{G}'}{f_{G}\vee\rho},\ldots,\frac{f_{G}^{\left(k\right)}}{f_{G}\vee\rho}\right)-p_{k}\left(\frac{f_{G}'}{f_{G}},\ldots,\frac{f_{G}^{\left(k\right)}}{f_{G}}\right)\right\Vert \\
 &  & +\left\Vert p_{k}\left(\frac{f_{\hat{G}}'}{f_{\hat{G}}\vee\rho},\ldots,\frac{f_{\hat{G}}^{\left(k\right)}}{f_{\hat{G}}\vee\rho}\right)-p_{k}\left(\frac{f_{\hat{G}}'}{f_{\hat{G}}},\ldots,\frac{f_{\hat{G}}^{\left(k\right)}}{f_{\hat{G}}}\right)\right\Vert \Biggr]
\end{eqnarray*}
\end{lem}
\begin{proof}
For any $\rho$
\begin{eqnarray*}
\left\Vert \hat{\kappa}_{k}\left(\param\given y\right)-\kappa_{k}\left(\param\given y\right)\right\Vert  & \leq & \left\Vert \hat{\kappa}_{k}\left(\param\given y\right)-\hat{\kappa}_{k,\rho}\left(\param\given y\right)\right\Vert +\left\Vert \hat{\kappa}_{k,\rho}\left(\param\given y\right)-\kappa_{k}\left(\param\given y\right)\right\Vert \\
 & = & \left\Vert p_{k}\left(\frac{f_{\hat{G}}'}{f_{\hat{G}}\vee\rho},\ldots,\frac{f_{\hat{G}}^{\left(k\right)}}{f_{\hat{G}}\vee\rho}\right)-p_{k}\left(\frac{f_{\hat{G}}'}{f_{\hat{G}}},\ldots,\frac{f_{\hat{G}}^{\left(k\right)}}{f_{\hat{G}}}\right)\right\Vert +\left\Vert \hat{\kappa}_{k,\rho}\left(\param\given y\right)-\kappa_{k}\left(\param\given y\right)\right\Vert .
\end{eqnarray*}
Applying Lemma \ref{lem:regularized-continuous} and taking the minimum
over $\rho$ finishes the proof.
\end{proof}
\bibliographystyle{plainnat}
\bibliography{21_Users_omuralidharan_Google_Drive_Projects_Emp_Bayes_CI_referencesDatabase}

\end{document}